\documentclass[sigconf,screen,10pt]{acmart}

\makeatletter
\def\@ACM@checkaffil{
    \if@ACM@instpresent\else
    \ClassWarningNoLine{\@classname}{No institution present for an affiliation}%
    \fi
    \if@ACM@citypresent\else
    \ClassWarningNoLine{\@classname}{No city present for an affiliation}%
    \fi
    \if@ACM@countrypresent\else
        \ClassWarningNoLine{\@classname}{No country present for an affiliation}%
    \fi
}
\makeatother

\usepackage[english]{babel}
\usepackage{blindtext}
\usepackage{amsmath, amssymb, amsthm}
\usepackage{pifont}
\newcommand{\xmark}{\ding{55}}%
\usepackage{balance}
\usepackage{colortbl}
\usepackage{tabularx}
\usepackage{multirow}
\usepackage{graphicx}
\usepackage{tikz} 
\usepackage{filecontents}
\usepackage{xspace}
\usepackage[titletoc]{appendix}
\usepackage{listings}
\usepackage{xcolor}
\usepackage{verbatim}
\usepackage[usestackEOL]{stackengine}
\usepackage[colorinlistoftodos]{todonotes}
\usepackage{hyperref}
\usepackage{url} 
\usepackage{times}  
\usepackage[linesnumbered]{algorithm2e} 
\usepackage{subcaption}
\usepackage{array}
\usepackage{soul}
\usepackage{textcomp}

\usepackage{caption}
\def\compactify{\itemsep=0pt \topsep=0pt \partopsep=0pt \parsep=0pt}
\let\latexusecounter=\usecounter

\newenvironment{CompactEnumerate}
  {\def\usecounter{\compactify\latexusecounter}
   \begin{enumerate}}
  {\end{enumerate}\let\usecounter=\latexusecounter}
\newcommand{\cut}[1]{}

\newcommand{\etc}{{etc.}\xspace}

\newcommand{\para}[1]{\vspace{1ex}\noindent\textbf{#1.}\xspace}
\newcommand{\Para}[1]{\vspace{1ex}\noindent\textbf{#1}\xspace}

\newcommand{\nop}[1]{}

\newcommand{\aarti}[1]{\textcolor{cyan}{[Aarti: #1]}}
\newcommand{\xiangyu}[1]{\textcolor{blue}{[Xiangyu: #1]}}

\newcommand{\sysname}{CaT\xspace}

\definecolor{Gray}{gray}{0.85}
\definecolor{LightCyan}{rgb}{0.88,1,1}

\renewcommand\footnotetextcopyrightpermission[1]{} 
\acmDOI{}

\acmISBN{}

\acmConference[]{}
\acmYear{}
\copyrightyear{}

\acmPrice{}

\begin{document}
\title{High-Level Synthesis for Packet-Processing
Pipelines}


\newcommand{\aut}[2]{#1\texorpdfstring{$^{#2}$}{(#2)}}
\author{\aut{Xiangyu Gao}{1}, \aut{Divya Raghunathan}{2}, \aut{Ruijie Fang}{2}, \aut{Tao Wang}{1}, \aut{Xiaotong Zhu}{1}, \aut{Anirudh Sivaraman}{1}, \aut{Srinivas Narayana}{3}, \aut{Aarti Gupta}{2}}
\affiliation{\aut{}{1}\textit{New York University}\quad
             \aut{}{2}\textit{Princeton University}\quad
             \aut{}{3}\textit{Rutgers University}}


\begin{abstract}
{\it
Compiling high-level programs to target high-speed packet-processing pipelines is a challenging combinatorial optimization problem. 
The compiler must configure the pipeline's resources to match the high-level semantics of the program, while packing all of the program's computation into the pipeline's limited resources.
State of the art approaches tackle individual aspects of this problem. Yet, they miss opportunities to efficiently produce globally high-quality outcomes. 

We argue that High-Level Synthesis (HLS), previously applied to ASIC/FPGA design, is the right framework to decompose the compilation problem for pipelines into smaller pieces with modular solutions.
We design an HLS-based compiler that works in three phases.
{\em Transformation} rewrites programs to use more abundant pipeline resources, avoiding scarce ones. 
{\em Synthesis} breaks complex transactional code into configurations of pipelined compute units. 
{\em Allocation} maps the program's compute and memory to the hardware resources.

We prototype these ideas in a compiler, \sysname, which targets the Tofino pipeline and a cycle-accurate simulator of a Verilog hardware model of an RMT pipeline.
\sysname can handle programs that existing compilers cannot currently run on pipelines, generating code faster than existing compilers, while using fewer pipeline resources. 

} 
\end{abstract}
\maketitle


\section{Introduction}

Reconfigurable packet-processing pipelines (RMT~\cite{rmt}) are emerging as important programmable platforms, found in high-speed network switches and network interface cards (NICs).
Examples include the Tofino~\cite{p4sde}, Trident~\cite{trident}, 
Jericho switches~\cite{jericho}, 
the Pensando~\cite{pensando}, and Intel IPU NICs~\cite{intel_ipu}.

Programmable pipelines are organized into multiple stages, where each stage processes one packet in parallel, and hands it off to the next stage (\S\ref{sec:pipelines-background}).
Each stage contains memory blocks to hold tables containing packet-matching rules and state (e.g., counters) maintained across packets. 
Header fields are extracted from packets to match the table rules.
Once the packet's fields are matched against a rule, the packet or state can also be updated 
using an action.

P4~\cite{p4_16} is emerging as a popular language to program these pipelines.
P4 offers the ability to parse packets according to custom header definitions, and specify the match types and actions on parsed packets.
A P4 action may modify packet header fields and state. 

\Para{The Compilation Problem.} The community has developed P4 solutions targeting programmable pipelines for many use cases, such as in-network computation~\cite{netcache}, monitoring~\cite{pint}, load balancing~\cite{silkroad}, and security~\cite{nethide}. 
A compiler translating P4 programs to high-speed pipelines must solve a hard combinatorial optimization problem with several aspects:

\noindent {\em (1) Resource balance:} There are multiple pipeline resources, with some resources being scarce (stages, gateways, \etc) and others being abundant (ALUs). 
Some resources must be allocated hand in hand (e.g., match memory and ALUs).

\noindent {\em (2) Semantics preservation:} P4 actions are transactional~\cite{p4-actions-spec}, executing to completion on each packet before processing the next one. If a user wishes to perform an algorithmic action on packet data that requires multiple stages, the compiler must be able to split the action into multiple ALUs and stages, ensuring the implementation respects the transactional semantics of the computation \cite{domino}. 

\noindent {\em (3) All-or-nothing fit:} A program targeting a high-speed pipeline will either run at the full pipeline rate, or cannot run at all. It is paramount to ``pack'' all of the P4 program into the pipeline's limited resources. 

Prior work has tackled several individual aspects of this compilation problem (\textsection\ref{sec:prior-compilers}). 
A piecemeal approach loses opportunities to globally reduce resource usage (e.g., stages), which is necessary to fit complex programs on the pipeline. 
However, it is intractable to solve a single combinatorial optimization problem. 
Our goal is to find the right decomposition of the large problem into smaller pieces, enabling global optimization of resource use with modular solution to each piece.

\Para{Our Approach.} 
In this paper, we present an end-to-end compiler, \sysname, that unifies prior approaches and translates high-level P4 programs into a low-level representation suitable for hardware execution. 
Our main idea is to adapt {\em high-level synthesis} (HLS)---a technology for improving productivity of 
hardware design for ASICs~\cite{hls_asic} and FPGAs~\cite{hls_fpga}---to the domain of packet-processing pipelines.

Informally, HLS~\cite{hls} takes as input a high-level algorithmic description of the hardware design with no reference to clocks or pipelining, and with limited parallelism in the description. 
An HLS compiler then progressively lowers this high-level description down to an optimized hardware implementation, pipelining the implementation if possible, executing multiple computations in parallel, and mapping computations down to a register-transfer level (RTL) design. 

We believe that adapting HLS for compilers targeting packet-processing pipelines will raise the user's level of programming abstraction, while retaining the performance of low-level pipeline programming.
For a user developing algorithmic programs in P4 (such as those used for in-network computation, e.g.,~\cite{siphash, netcache, daiet-hotnets17}), HLS techniques eliminate the labor of manually breaking the high-level algorithmic computation into actions spread over many pipeline stages (\S\ref{sec:case-for-hls}). 
At the same time, HLS techniques offer many distinct advantages for targeting packet-processing pipelines (\S\ref{sec:case-for-hls}).


\begin{figure}[!t]
    \centering
    \includegraphics[width = 
    \columnwidth]{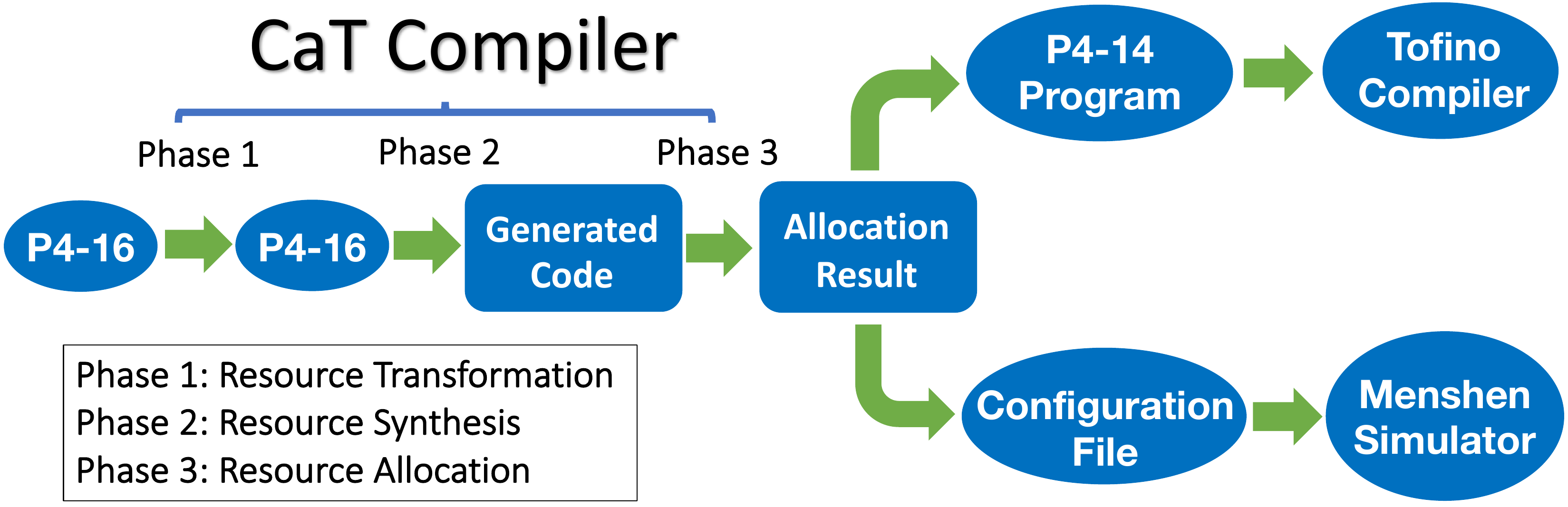}
    \begin{tiny}
    \caption{The workflow of CaT compiler.}
    \label{fig:hls_workflow}
    \end{tiny}
\end{figure}

The workflow of our compiler, \sysname, is shown in Figure~\ref{fig:hls_workflow}. It consists of three phases that roughly mirror a traditional HLS compiler. 
The input consists of P4 code that contains tables with matches and action code blocks manipulating packet headers and state.
The action code blocks may be written without regard to its feasibility in a single pipeline stage.
The first phase of \sysname employs \emph{resource transformations} that rewrite a high-level P4 program to another
semantically-equivalent 
high-level P4 program; these rewrites are used to transform a computation's use of one scarce resource to its use of a relatively abundant resource, and potentially reduce the number of stages as well. 
The second phase performs \emph{resource synthesis} to lower transactional blocks of statements in the high-level P4 program to a lower-level program suitable for hardware execution. 
In this step, individual 
ALUs in hardware 
are configured to realize the programmers' intent in the transactional action blocks, while respecting their computational capabilities. 
The third phase performs \emph{resource allocation} to allocate the computation units corresponding to the lowered 
program to physical resources such as ALUs and memory in the pipeline.
Notably, our HLS workflow works within the confines of the widely used P4 ecosystem without requiring the development of a new DSL for packet processing.


\Para{Our Contributions.} 
The main technical contribution of \sysname's three-phase approach is the modularization of the large combinatorial optimization problem of compilation into smaller problems, whose solutions still enable a high-quality global result (\S\ref{sec:design}).
Additionally, we improve upon the state of the art and introduce new techniques in each phase. 
In particular, our resource transformations (\S\ref{ssec:rewrite}) are driven by a novel \emph{guarded dependency analysis} that identifies false dependencies, thereby exposing more parallelism opportunities when performing rewrites. 
Our resource synthesis phase (\S\ref{ssec:resource-synthesis}) uses a novel synthesis procedure that quickly finds pipelined solutions with good-quality results
for complex actions.
It separates out stateful updates from stateless updates, to decompose a large program synthesis problem into smaller and more tractable subproblems.
Stateless code is synthesized into a minimum-depth computation tree, 
i.e., with the minimum number of stages.
In comparison to prior work~\cite{chipmunk}, this new synthesis algorithm allows \sysname to handle many large actions, in a much shorter time, and with far fewer computational resources needed for compilation.
Finally, our resource allocation phase (\S\ref{ssec:resource_allocation}) uses a constraint-based formulation that extends prior work~\cite{lavanya_nsdi15} to handle complex multi-stage transactional actions. 
Our techniques can support general P4 programs (including @atomic constructs) efficiently, including programs translated into P4 from higher-level domain-specific languages developed for pipeline programming~\cite{domino, chipmunk, lyra, lucid, p4all}.

Our prototype of \sysname can target: (1) the Tofino pipeline, and (2) an open-source RMT pipeline called Menshen (that was implemented into an FPGA)~\cite{menshen,menshen_site}.
Existing commercial switches have proprietary instruction sets that preclude the kind of low-level resource allocation and control over ALU configurations implemented by \sysname. 
Therefore, our backend for Tofino~\cite{p4sde} 
generates low-level P4 in lieu of machine code. 
To evaluate \sysname in full generality, we extend Menshen's
open-source register-transfer level (RTL) Verilog model with additional resources for our experiments.
We generate code for the cycle-accurate simulator of Menshen, 
and 
also use it for testing the CaT prototype. 
Our results (\S\ref{sec:eval}) show that CaT can automatically compile programs that require manual changes to be accepted by the Tofino compiler. On other challenging benchmarks, CaT produces close-to-optimal code 
and does so about 3 times faster (on average) 
than prior work~\cite{chipmunk}.

\vspace*{-0.1in}
\section{Background and Related Work}

\subsection{Packet-Processing Pipelines}
\label{sec:pipelines-background}

The compiler target in this paper is a programmable packet-processing pipeline following the Reconfigurable Match-Action Tables (RMT) architecture~\cite{rmt}. 
Such pipelines are present in commercially available programmable switches such as the Barefoot Tofino~\cite{p4sde}, Broadcom Trident, and Mellanox Spectrum, and NICs such as the Pensando DPU. 
An RMT-style pipeline consists of (i) a programmable packet parser, (ii) a number of processing stages structured around match-action computation. We describe these components briefly below.

A programmable parser takes in a user-specified parsing grammar, and extracts packet header fields.
These set of fields, termed the {\em packet header vector}, are permitted to be both read and written in each pipeline stage.
One match-action stage extracts relevant fields from the PHVs using a crossbar circuit.
The fields are then matched against user-inserted rules in stage-local match memory.
The memory may also contain {\em state}, i.e., values maintained on the switch and updated by every packet, such as a packet counter.
Once a packet matches a rule, a corresponding set of actions is invoked.
The actions are implemented using Very Long Instruction Word (VLIW) ALUs which may modify multiple PHV fields in one shot.
Some match-action tables may be skipped entirely (e.g., due to control flow) through 
circuit components called {\em gateways.}

Two factors limit the available resources and expressiveness of packet-processing pipelines.
First, to support high throughput (e.g., 6 Tbit/s in Tofino), pipelines are clocked at high frequencies (e.g., 1 GHz for Tofino). 
Second, the pipeline must admit a new packet every clock cycle.
Hence, stateful computations (read-modify-write) must finish in one clock cycle.
Further, stage-local memories are limited in density and size, to support fast lookup.
Finally, constraints on chip area and power limit the number of pipeline stages (e.g., 12 match-action stages in Tofino) and control circuitry (e.g., number of gateways and crossbars).

Such exacting hardware constraints would challenge any compiler.
Program behavior is {\em all-or-nothing}: a program that fits into the pipeline resources would run at clock rate, otherwise it simply cannot be run.
There is no graceful degradation between these extremes.

\subsection{Related Work}
\label{sec:prior-compilers}

There has been significant interest in developing compilers and domain-specific languages (DSLs) for packet-processing pipelines. We categorize the existing compiler efforts based on their support for: program rewriting, code generation, and resource allocation. 

\para{DSLs for programming packet pipelines}
P4 and NPL are the most widely used languages to program packet pipelines. They share many syntactic and semantic aspects. 
Several academic projects have proposed new DSLs or extensions to P4 to remedy many of P4's shortcomings. 
For instance, microP4~\cite{microp4} adds modularity to P4. 
Lyra~\cite{lyra} addresses the issue of portability of programs across multiple devices. 
Lyra and FlightPlan~\cite{flightplan} address the problem of partitioning a program automatically across multiple devices. Lucid~\cite{lucid} introduces an event-driven programming model for control applications in the data plane.
P4All~\cite{p4all} extends P4 to support `elastic' data structures, whose size can grow and shrink dynamically based on the availability of switch resources. 
Domino~\cite{domino} is a DSL that supports transactional packet processing: a programmer specifies a block of code that is executed on each packet in isolation from other packets. 
These languages can all be translated into P4, and in this paper, we directly take P4 programs as our starting point. This provides opportunities to explore complementary combinations of their techniques. 
One limitation is that CaT does not currently handle the problem of partitioning a \emph{network-wide} program into per-device programs, like Lyra and Flightplan. Instead, our goal is to build a high quality compiler that inputs a P4 program for a \emph{single device} and outputs a high-quality implementation for that device. 

\para{Program rewriting} 
The open-source reference P4 compiler~\cite{p4c}, which is the foundation for most P4 compilers including the widely-used Tofino compiler~\cite{p4sde}, employs rewrite rules to turn an input P4 program into successively simpler P4 programs. 
These rewrite rules consist of classical optimizations like common sub-expression elimination and constant folding. 
Rewrite rules are also employed by Cetus~\cite{cetus} and Lyra~\cite{lyra} to merge tables in different stages (under certain conditions) into a single ``cartesian-product'' table in a single stage, thereby saving on the number of stages.
\sysname uses rewrite rules to transform uses of scarce resources (gateways, stages) to more abundant ones (tables, memory, ALUs). 

\para{Code generation for complex actions}
Domino~\cite{domino} and Chipmunk~\cite{chipmunk} tackle the problem of {\em code generation}, i.e., selecting the right instructions (that configure ALU opcodes) for a program action expressed in a high-level language.
These compilers 
have to respect the limited capabilities of each stage's VLIW ALUs while correctly implementing state updates according to @atomic semantics for transactions (\S\ref{sec:pipelines-background}). 
Domino largely uses rewrite rules and employs program synthesis to code-generate just the stateful fragments in the action, 
but it may fail  
on some semantically-equivalent programs. 
Chipmunk employs program synthesis to exhaustively search for ALU configurations 
for any semantically-equivalent program, but
at the expense of high compile time.
Lyra~\cite{lyra} uses {\em predicate blocks,} chunks of code predicated by the same path condition, to break up algorithmic code into smaller blocks that have only inter-block (but no intra-block) dependencies. 
\sysname's resource synthesis is faster than Chipmunk's and more reliable than Domino's (Table~\ref{table:Code_generation_time_comparison_domino_tofino}, \S\ref{ssec:eval_synthesis}). It generalizes Lyra's predicate block approach by considering ALUs expressed via a parameterizable grammar, such that the procedure is independent of the operations in the program's source code or intermediate representation.

\para{Resource allocation}
The problem of allocating specific resources required by a P4 program (e.g., match memory blocks, a specific number of ALUs, etc.) can be posed as an integer linear programming problem (ILP)~\cite{lavanya_nsdi15, p4all} or as a constraint problem~\cite{lyra} for Satisfiability Modulo Theory (SMT) solvers~\cite{smt-hb}.
If the constraints of the hardware are modeled precisely, ILP-based techniques can improve resource allocation relative to greedy heuristics for resource allocation.
\sysname's resource allocation (\S\ref{ssec:resource_allocation}) indeed uses a fine-grained constraint-based formulation that models detailed pipeline resources and enables global optimization by considering dependencies across tables as well as within actions.

\definecolor{BadC}{RGB}{255,204,204}
\definecolor{GoodC}{RGB}{204,255,204}
\definecolor{FairC}{RGB}{220,220,220}
\begin{table*}[]
    \tiny
    \centering
    \begin{scriptsize}
    \resizebox{\textwidth}{!}{  
    \begin{tabular}{lcccccc}
         \toprule
         \textbf{Project} & \textbf{Program Rewriting} & \textbf{Code Generation} & \textbf{Resource Allocation} & \multicolumn{2}{c}{\textbf{Retargetability}} & \textbf{Language Constructs} \\
         & & & & \textbf{Instruction Sets} & \textbf{Resource Constraints} & \\ 
         \midrule
         Domino~\cite{domino} & \cellcolor{GoodC}Yes & \cellcolor{GoodC}Rewriting, program synthesis & \cellcolor{BadC}{\xmark} & \cellcolor{GoodC}Atom templates & \cellcolor{BadC}{\xmark} & \cellcolor{GoodC}Packet transactions \\
         Chipmunk~\cite{chipmunk} & \cellcolor{BadC}\xmark & \cellcolor{GoodC}Program synthesis & \cellcolor{BadC}\xmark &  \cellcolor{GoodC}ALU DSL & \cellcolor{BadC}{\xmark} & \cellcolor{GoodC}Packet transactions\\
         Lyra~\cite{lyra} & \cellcolor{GoodC}Yes & \cellcolor{BadC}\xmark & \cellcolor{GoodC}SMT & \cellcolor{BadC}{\xmark} & \cellcolor{GoodC}SMT constraints & \cellcolor{GoodC}Network-wide programs \\
         Flightplan~\cite{flightplan} & \cellcolor{BadC}\xmark & \cellcolor{GoodC}Resource rules & \cellcolor{GoodC}Resource rules & \cellcolor{BadC}{\xmark} & \cellcolor{GoodC}Resource rules & \cellcolor{GoodC}Network-wide programs \\
         Cetus~\cite{cetus} & \cellcolor{GoodC}Yes & \cellcolor{BadC}\xmark & \cellcolor{GoodC}Table Merging, PHV Sharing & \cellcolor{BadC}\xmark & \cellcolor{GoodC}SMT constraints & \cellcolor{BadC}\xmark \\
         P4All~\cite{p4all} & \cellcolor{BadC}\xmark & \cellcolor{BadC}\xmark & \cellcolor{GoodC}ILP & \cellcolor{BadC}\xmark & \cellcolor{GoodC}ILP constraints & \cellcolor{GoodC}Elastic data structures \\
         Jose et al.~\cite{lavanya_nsdi15} & \cellcolor{BadC}\xmark & \cellcolor{BadC}\xmark & \cellcolor{GoodC}ILP & \cellcolor{BadC}\xmark & \cellcolor{GoodC}ILP constraints & \cellcolor{BadC}\xmark \\
         Lucid~\cite{lucid} &\cellcolor{BadC} \xmark & \cellcolor{GoodC}memops & \cellcolor{BadC}\xmark &  \cellcolor{GoodC}memops & \cellcolor{BadC}\xmark & \cellcolor{GoodC}Event-driven programming \\
         Tofino compiler~\cite{p4sde} & \cellcolor{GoodC}Yes & \cellcolor{GoodC}Yes & \cellcolor{GoodC}Heuristics & \cellcolor{BadC}\xmark & \cellcolor{BadC}\xmark & \cellcolor{BadC}\xmark \\
         \sysname (this work) & \cellcolor{GoodC}Yes & \cellcolor{GoodC}Min-depth tree algorithm & \cellcolor{GoodC}ILP/SMT &\cellcolor{GoodC}ALU grammar & \cellcolor{GoodC}ILP constraints & \cellcolor{GoodC}P4's atomic construct \\
         \bottomrule
    \end{tabular}}
    \caption{\sysname unifies prior work on program rewriting, code generation, resource allocation, and does so within the context of the P4 language, without needing a new DSL.\label{tab:priorWorks}}
    \end{scriptsize}
\end{table*}
%


\begin{table*}[h]
\tiny
\resizebox{\textwidth}{!}{  
\begin{tabular}{p{3cm}p{3.25cm}p{3cm}p{3.6cm}p{3.2cm}}
\toprule
\textbf{CaT compiler phase} & \textbf{Technique} & \textbf{Builds on prior work} & \textbf{Differences in CaT} & \textbf{Other complementary work}  \\
\midrule
\textbf{1: Resource transformation} (corresponds to HLS code transformation) & Rewrite rules & LLVM~\cite{llvm}, HLS~\cite{hls, hls_fpga}, p4c~\cite{p4c} & Rewrite rules target RMT, based on novel guarded dependency analysis & p4c~\cite{p4c} uses platform-independent rewrites, Cetus~\cite{cetus} merges tables \\ \midrule
\multirow{4}{3cm}{\textbf{2: Resource synthesis} (corresponds to HLS operation binding and HLS module selection)} & Preprocessing: branch removal, SSA, SCC for stateful updates & Domino~\cite{domino}, SSA~\cite{ssa}, VLIW~\cite{lam} & No backward control flow (like Domino) & \\ 
& Simplifications: const prop, expr simplification, deadcode elimination & LLVM~\cite{llvm}, HLS code transformation~\cite{hls,hls_fpga} & No backward control flow & \\ 
& Uses program synthesis for mapping operations to ALUs & Chipmunk~\cite{chipmunk} & Novel synthesis procedure: faster, more scalable, uses smaller queries  & Lucid~\cite{lucid} uses syntactic rules to ensure operations map to Tofino \\
& Target portability via parameterizable grammars for ALUs & Sketch~\cite{sketch}, Chipmunk~\cite{chipmunk} & Generate resource graph (used in Phase 3), not low-level ALU configs & \\ \midrule
\multirow{4}{3cm}{\textbf{3: Resource allocation} (corresponds to HLS scheduling)} & Constraints for match memories & Jose et al.~\cite{lavanya_nsdi15}, Lyra~\cite{lyra} & Associates match memories with corresponding action resources & \\ 
& Constraints for multi-stage actions & Domino~\cite{domino}, Chipmunk~\cite{chipmunk} & Uses result of Phase 2 for intra-action dependencies and ALU output propagation & \\ 
& Constraints for multiple transactions & Domino~\cite{domino}, Chipmunk~\cite{chipmunk} handle a single transaction only & Enforces inter-table and intra-action dependencies for global optimization & \\
& Modeling real hardware constraints in backend FPGA target & Menshen provides a FPGA backend target~\cite{menshen} & Extended functionality of resources available in Menshen & Tofino compiler uses heuristics for resource allocation \\ \bottomrule
\end{tabular}}
\caption{CaT in relation to HLS and other related work.}\label{tab:hls}
\end{table*}

\subsection{Why HLS?}
\label{sec:case-for-hls}
\sysname is an end-to-end compiler for P4-16 programs that takes inspiration from HLS to provide both: (1) a high-level of abstraction for specifying packet-processing functionality, and (2) high quality of the compiler-produced implementation. While prior approaches to HLS for ASICs and FPGAs have sometimes resulted in poor quality of the generated implementation, we believe that the narrower domain of packet-processing pipelines 
is particularly well-suited for applying HLS more gainfully than has been done before. 
First, HLS techniques are designed to systematically explore tradeoffs between the functionality (e.g., which ALU can implement an operation?), capacity (e.g., how many ALUs, gateways, etc.?), and scheduling (which stage should run an operation?) of resources--- a core challenge in compiling to packet-processing pipelines.
Second, HLS techniques can be effective in pipelining code with memory state updates, to enable transactional semantics, an abstraction that provides 
the illusion that all previous packets have finished their computation over the state.
Importantly, an HLS workflow can combine these ideas with code rewrite and program synthesis.

\begin{figure}[t]
    \centering
    \includegraphics[width=0.35\textwidth]{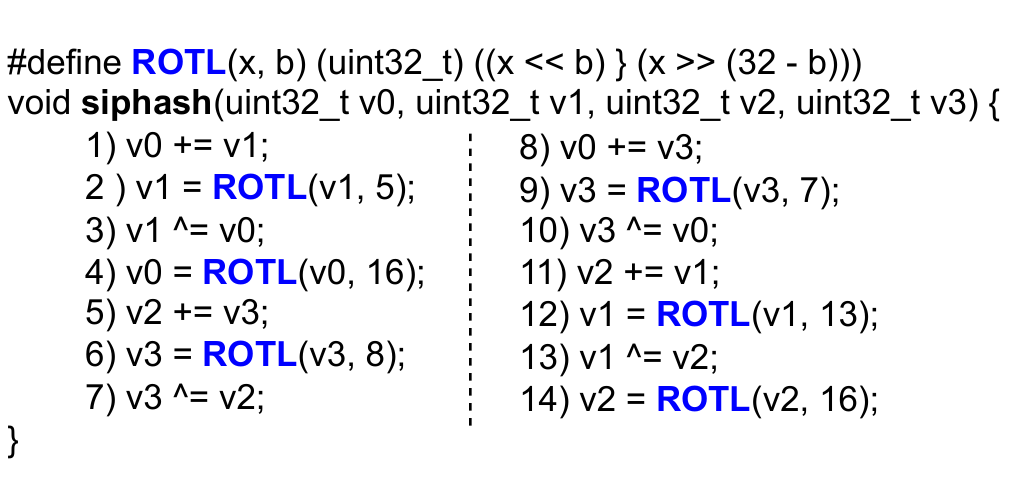}
    \caption{Motivating Example ME-1: SipHash was manually split into four stages and rewritten by P4 users~\cite{siphash}.}
    \label{fig:motivation_siphash}
\end{figure}

Today, developers write down actions in P4 programs with the assumption that each action must finish in one stage.
However, tracking the hardware-level feasibility of an action leads to an unnecessarily low level of abstraction, especially when developing high-speed algorithmic code.
Consider the example pseudocode (motivating example ME-1) shown in Figure~\ref{fig:motivation_siphash}. 
This function implements the SipHash algorithm, used as a hash function to prevent collision-based flooding attacks~\cite{siphash-indocrypt12}.
The developer of a P4-version of this algorithm (distinct from the authors of this paper) started with a high-level description of the algorithm (Table 3, ~\cite{siphash}).
The developer then {\em manually} changed it into a pipelined implementation (Table 4,~\cite{siphash}), because the algorithm as expressed cannot be compiled by the Tofino compiler since it cannot be finished in one stage.
We argue that a good compiler should automate this. 
Indeed, \sysname can successfully handle this example (discussed in \S\ref{ssec:eval_synthesis}), without requiring an expert developer to manually pipeline their code.


To produce high quality implementations, \sysname combines ideas from several prior projects (Table~\ref{tab:priorWorks}) into an end-to-end system for the first time. 
The HLS workflow of \sysname divides up the process of compiling a P4-16 program into three phases -- their correspondence with traditional HLS steps~\cite{hls} and relation with prior work are highlighted in Table~\ref{tab:hls}.
As shown, our resource transformation corresponds to the code transformation step in HLS, our resource synthesis corresponds to operation binding (or module selection~\cite{hls_fpga}) in HLS, and our resource allocation corresponds to the scheduling step in HLS. 
To further improve the effectiveness of HLS in our setting, we customize these phases to RMT pipelines. Specifically, our resource transformations focus on scarce resources (pipeline stages, gateways). Our resource synthesis procedure leverages the strict requirement that stateful updates must fit in a single stage to ease the task of pipelining, which is often challenging for HLS in general.
Our resource allocation phase uses the result of synthesis on action blocks, to pack multiple action blocks in each pipeline stage.
Thus, while these phases build on top of well-known prior work including HLS efforts (Column~3), there are significant differences and novelties in our work (Column~4). The last column lists complementary techniques that could be combined in our compiler. 

\section{Compiler Design}
\label{sec:design}

The general problem of optimal code generation in compilers is known to be \textsf{NP}-complete~\cite{AJU77}, and most compilers decompose the problem in some way to tradeoff optimality for reasonable performance. 
In our HLS-based approach, Phases 2 and 3 can be viewed as an \emph{action-level modularization} of the overall compilation problem, where we perform \emph{local} resource synthesis for each action block in the P4 program, and then use these results to perform a \emph{global} resource allocation for all actions blocks. This keeps the synthesis runtime manageable in practice while providing good-quality results on resource usage.  
Note that we support rich computations in actions that could require multiple stages as well as transactional (@atomic) semantics. In the remainder of the paper, we refer to such rich action-computations as \emph{transactions}. We now describe the three phases of our compiler in detail.


\subsection{Phase 1: Resource Transformation}
\label{ssec:rewrite}
In the first phase of our compiler, we perform source-to-source rewrites in P4, with the goal of transforming a program that makes use of scarce resources, to one that makes use of more abundant resources. Rewrite rules provide a flexible and general approach for this purpose, and can be easily extended by adding more rules for new backend targets and resources. 
\sysname includes rewrite rules for if-else statements in the control block of a P4 program. 
The standard p4c compiler transforms each action in an if-else branch into one default table, i.e., a table without a match key and with only one action. Our rewrites effectively transform multiple (possibly nested) if-else statements into one bigger table with keys, thereby using less gateway resources than multiple default tables dispersed in if-else branches. 
More importantly, our rewrites are driven by a novel \emph{guarded dependency analysis} that helps identify parallelism opportunities by eliminating false dependencies -- this often results in reducing the number of stages as well.


\begin{figure}[t]
    \centering
    \includegraphics[width=0.35\textwidth]{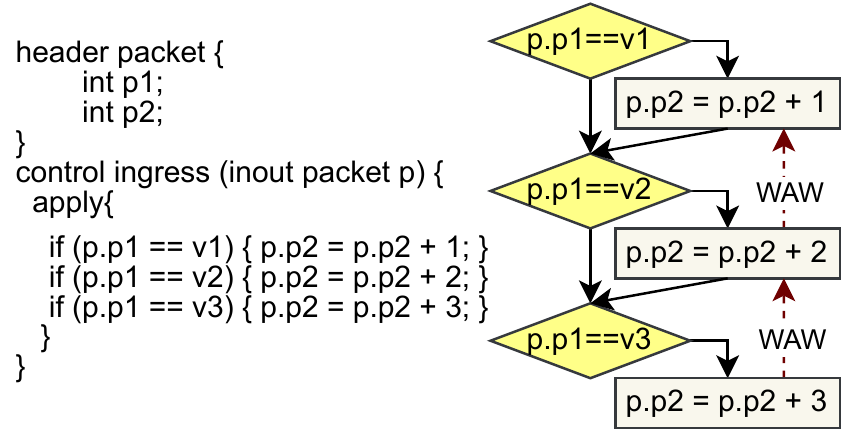}
    \caption{Motivating Example ME-2: The control flow graph of a P4 control block (left), simplified from SipHash~\cite{siphash}. Dependencies shown as dotted red edges. $v1\neq v2 \neq v3$ are constants.}
    \label{fig:p4cfgcb}
\end{figure}

\subsubsection{Guarded dependency analysis.}
The sequence of program statements inside the \texttt{apply \{...\}} block of a P4 control block can be treated as a
well-structured branching program (without loops) with (possibly nested) if-statements, reads and writes to PHV fields, and \texttt{apply} statements, which apply match-action tables. 
This program induces Read-after-Write (RAW), Write-after-Read (WAR), and Writer-after-Write (WAW) dependencies between pairs of program statements. 
These dependencies 
must be respected during synthesis and resource allocation. 
Conventionally, these dependencies are defined between pairs of program statements without accounting for \emph{path conditions}~\cite{symbex}, i.e., conditions under which a control path in a program is executed.
Specifically, a dependency pair 
denoted as $(v@s_1\to v@s_2, t)$ 
represents a $t$-dependency, $t\in \{RAW,WAR,WAW\}$, due to variable $v$ between statements $s_1$ and $s_2$. 

Consider the motivating example ME-2 shown in Figure~\ref{fig:p4cfgcb}, also inspired by a part of a real SipHash P4 program developed by other users~\cite{siphash}. The WAW dependencies (shown on the right) lead to requiring 3 pipeline stages. However, \emph{these WAW dependencies are not real}, since the if-conditions guarding these assignments are disjoint. 
Indeed, a developer of this program recognized the disjoint conditions and \emph{manually changed} the program to use nested-if statements, which requires 1 stage. 
We would like to \emph{automate} such rewrites. 
In particular, p4c and the Tofino compiler miss these rewrites in ME-2, likely due to a conservative dependency analysis.


To solve this issue, we propose
\emph{guarded dependencies}, which take into account path conditions along control paths.
Given a control-flow graph (CFG) $C$ for a P4 control block, a \emph{guarded dependency} between nodes $(n_1, n_2)\in C$ is defined as a tuple $(v@s_1\to v@s_2, t, \phi)$, where $v$ is the variable of concern at statement $s_1$ (in node $n_1$) and statement $s_2$ (in node $n_2$),  $t\in\{RAW,WAR,WAW\}$, and $\phi$ (called a guard) is a formula that describes all the path conditions under which node $n_2$ may be visited after node $n_1$ is visited. 
A procedure based on symbolic execution~\cite{symbex} or model checking~\cite{mc-hbmc} that computes path conditions can be used to determine precise guarded dependencies for the program.
In particular, we can use an SMT solver to identify \emph{false dependencies}, i.e., dependencies where $\phi$ is unsatisfiable. 
Since our goal is to enable rewrite rules for resource transformation, we perform a simpler lightweight static analysis described next.
\begin{figure}[t]
    \centering
    \includegraphics[width=\linewidth]{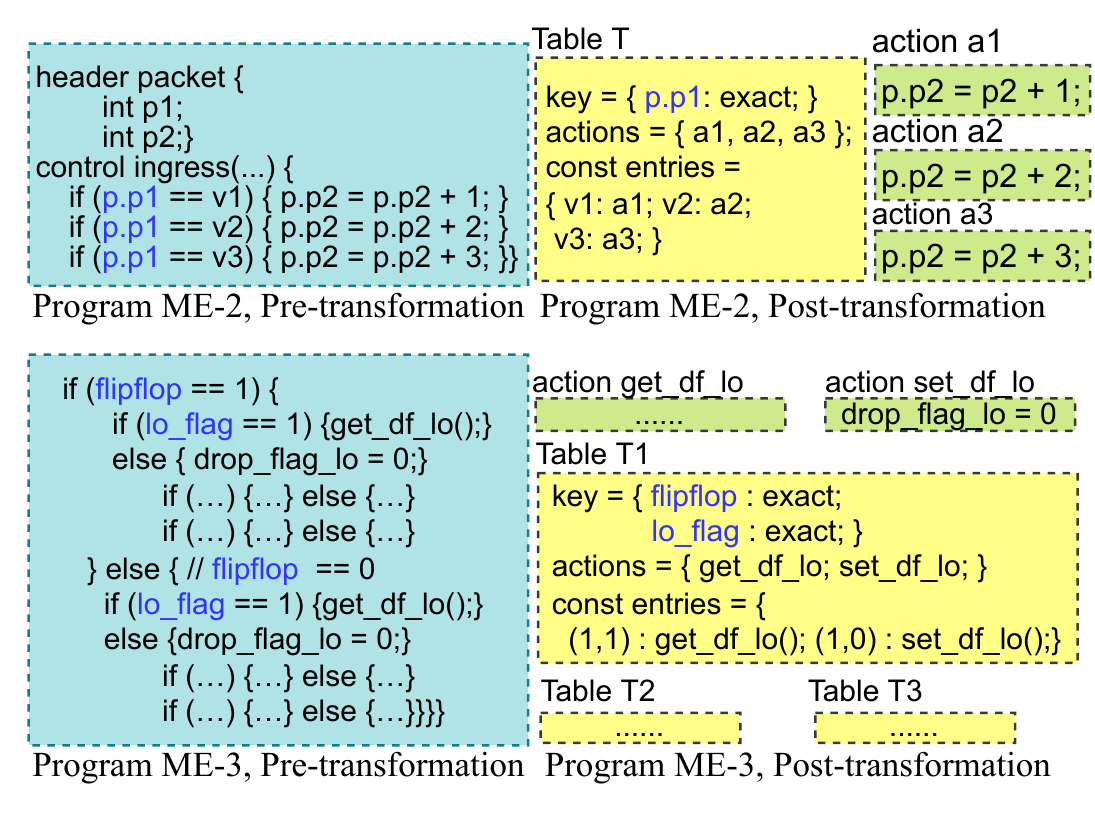}
    \caption{Illustration of Phase 1 Rewrites on motivating examples ME-2 (from Figure \ref{fig:p4cfgcb}), 
    and ME-3 (taken from a UPF Rate\_enforcer example provided by P4 users).}
    \label{fig:p4Canonical}
\end{figure}
\subsubsection{Lightweight guarded dependency analysis to support CaT rewrites.} First, we check that none of the assignment statements update any variables used in conditions of if-else statements. Such updates lead to RAW dependencies and more complicated analysis for rewrites -- we currently choose to not perform any rewrites in such cases. 
When there are no such updates, the path condition for each CFG node is a simple conjunction of branch conditions, 
which we compute by a depth-first traversal of the CFG, where branch conditions are pushed/popped on a stack at branch/merge points, respectively. 
For a pair of nodes $(n_1, n_2)\in C$ with a guarded dependency, the guard $\phi$ is a conjunction of the computed path conditions for $n_1$ and $n_2$. If $\phi$ is unsatisfiable, then this is a false dependency and removed, otherwise it is conservatively retained as a dependency.
For our ME-2 example in Figure~\ref{fig:p4cfgcb}, this analysis finds that the shown WAW dependencies are false, and removes them. 

\subsubsection{Rewrites to Match-Action Tables.} 
We now focus on (possibly nested) if-else statements where the branch conditions are tests on packet fields that can be implemented as keys in a match-action table. Based on the guarded dependency analysis, if there is no dependency between the branches, then we can rewrite them into a match-action table. 
The key of the generated table is comprised from packet fields used in the if-else conditions, and the actions are the computations within each branch. 
For example, Figure~\ref{fig:p4Canonical} illustrates our rewrites on two P4 programs -- ME-2, and another motivating example ME-3 taken from a UPF Rate\_enforcer example (provided by others). 
After rewriting, both ME-2 and ME-3 use only match-action tables and thus no gateway resources. ME-2 uses only 1 stage post-rewriting vs. 3 stages before rewriting (due to false WAW dependencies). ME-3 also uses only 1 stage post-rewriting vs. 2 stages before rewriting (due to needing too many gateway resources to fit into 1 stage). 
These motivating examples drawn from real-world P4 programs show the effectiveness of our approach, where \emph{manual steps taken by a programmer to reduce resource usage are successfully automated by CaT}.  

\subsection{Phase 2: Resource Synthesis}
\label{ssec:resource-synthesis}
For the second phase, we propose a novel procedure to perform resource synthesis on each P4 action block. 
Like Chipmunk~\cite{chipmunk}, we too use the SKETCH~\cite{sketch} program synthesis tool to generate \emph{semantically equivalent} code where operations are mapped to hardware ALUs. 
However, there are two important differences from Chipmunk.
First, we aim to reduce synthesis time by creating \emph{smaller} synthesis queries for SKETCH, rather than invoking it for synthesis of the entire transaction.
Second, rather than creating low-level holes in the ALU grid architecture that are filled by SKETCH (as done by Chipmunk), our synthesis queries are \emph{parameterized by a grammar} that specifies hardware ALUs. 
These parameterizable grammars enable the same synthesis procedure to be used for different hardware backends, thereby improving compiler retargetability. 
Our compiler currently supports three different grammars: Tofino ALUs~\cite{p4sde}, Banzai ALUs~\cite{domino}, and ALUs for our simulation of Menshen~\cite{menshen}; more can be supported as needed. 
We now describe details of resource synthesis.
\subsubsection{Preprocessing of a P4 action block.} 
We preprocess each action block of the P4 program to prepare for synthesis.
We first use some standard preprocessing steps (similar to Domino~\cite{domino}), including (a) branch removal (by replacing assignments under branches to conditional assignments), 
(b) creating two temporary packet fields for each stateful variable -- \emph{pre-state field} (denoting its value before update) and \emph{post-state field} (denoting its value after update), and (c) conversion to static single-assignment (SSA) form~\cite{ssa}. 


In addition (and differently from Domino), we perform several static analyses during preprocessing: constant folding, expression simplification, and dead code elimination. These analyses are useful in simplifying the action block of a P4 program, thereby reducing the difficulty of the subsequent SKETCH queries. 
We neither add nor delete stateful variables during preprocessing simplifications, since it is unsound.


\subsubsection{Computation graph construction}
After preprocessing, we construct a dependency graph (similar to Domino),
with nodes for each program statement (a conditional assignment) and an edge for each RAW dependency.\footnote{Conversion to SSA form removes WAW and WAR dependencies.} Edges are also added between the pre/post-state fields of each stateful variable. The strongly connected components (SCCs) of this graph correspond to stateful updates, which are condensed to form a \emph{computation graph} $G$. Thus, $G$ is a directed acyclic graph (DAG) with nodes for program computations (some with stateful updates) and edges for RAW dependencies. Nodes in $G$ are partitioned into two sets: \emph{stateful nodes} are formed from SCCs on the dependency graph, containing a set of program statements that describe an atomic stateful update; \emph{stateless nodes} are 
the other nodes in the dependency graph. 
Each edge $(u, v)$ is mapped to a packet field variable that appears in the LHS of the assignment at $u$ and in the RHS of the assignment at $v$. 
We call sources of $G$ (excluding stateful variables) \emph{primary inputs (PIs)} -- these represent input packet fields. 
Inputs to stateful nodes and outputs of sinks of $G$ that are not temporaries (such as pre/post-state fields, created during preprocessing) are called \emph{primary outputs (POs)} -- these represent the final values written to packet fields. 
\subsubsection{Synthesis Procedure for a P4 Action}
\label{ssec:codegen}
Synthesis for a P4 action is now performed on the computation graph $G$. 
Our procedure, detailed below, consists of four main steps: 1) normalization; 2) folding and predecessor packing optimizations; 3) synthesis of stateful updates; 4) synthesis of minimum-depth solutions for stateless code. Pseudocode for the procedure is shown in Algorithm~1 in Appendix~\ref{appendix:alg1}. 

One main novelty of our synthesis procedure is that 
it \emph{decomposes the overall problem into multiple SKETCH queries}, many of which are much simpler than a query for an entire transaction. In particular, Step 3 checks that each stateful update fits into a single stateful ALU. If any such query fails, then we terminate the procedure and provide feedback to the user about the reason for failure. We also use small queries to perform optimizations in Step 2, to help Step 3 succeed. Finally, Step 4 uses queries to perform synthesis for computations that provide inputs to the stateful nodes and the POs in $G$.
Another novelty is that Step 4 synthesizes \emph{solutions of minimum-depth (i.e., minimum number of pipeline stages)} for the stateless code in $G$, by using SKETCH in an iterative loop. This enables CaT to handle rich stateless computations in multi-stage actions, while searching the space over all possible equivalent programs. 
Although Chipmunk can also handle multi-stage actions and an exhaustive search space, its queries consider an entire transaction, 
which makes it much slower in comparison (\S\ref{ssec:eval_synthesis}).
\para{Step 1: Normalization of computation graph}
\label{para:codegen_step1}
In the typical hardware backends that we target (e.g., Menshen, Tofino), 
a stateful ALU can output a single value that is either the pre-state or the post-state value of one of its stateful registers. 
In this step, we normalize $G$ to a graph such that  
each stateful node has only one output, and each  packet field labelled as an out-edge from a stateful node is either the \emph{pre-state field} or the \emph{post-state field}. 
\begin{figure*} [t]
    \centering
    \includegraphics[width=0.95\linewidth]{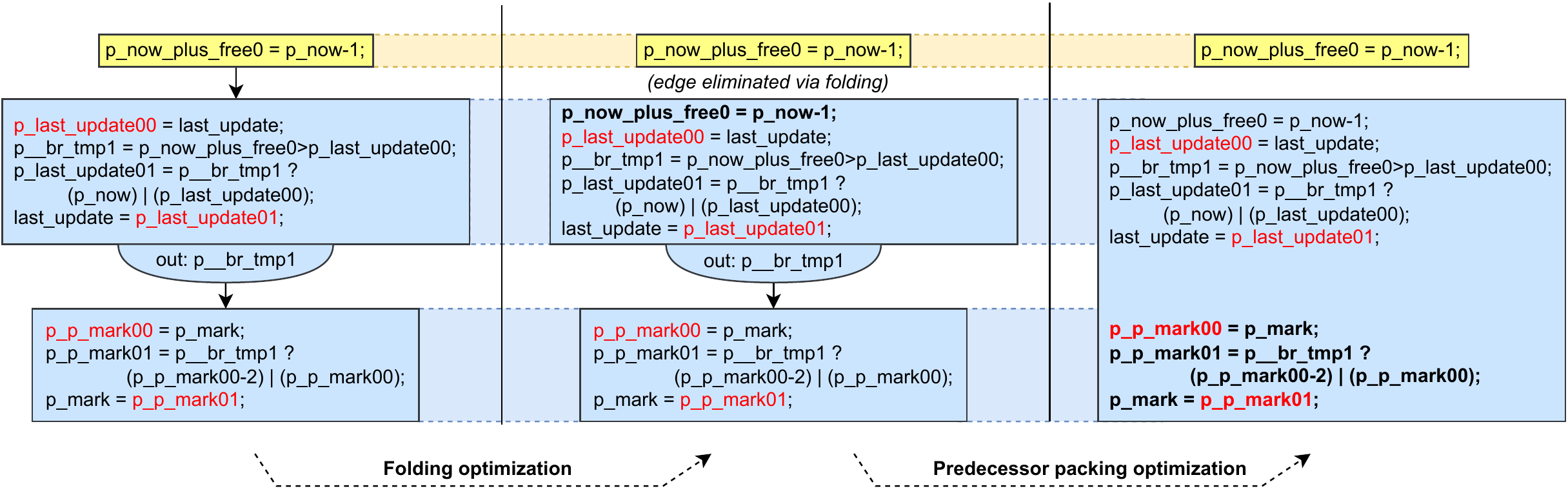}
   \caption{Computation graph for BLUE(decrease)~\protect\cite{blue} and sequence of optimizations performed when targeting the Tofino ALU. Stateful nodes shown in blue, stateless nodes shown in yellow, pre/post-state fields shown in red, modified parts shown in bold.}
    \label{fig:blue_decrease}
\end{figure*}

\para{Step 2: Folding and predecessor packing optimizations} 
We iterate the following two optimizations until convergence.

\emph{Folding to reduce input edges.} 
A stateful node with too many in-edges could cause Step 3 to fail, due to a limited number of inputs available in ALUs. The \emph{folding} optimization finds opportunities to reduce the number of in-edges to a stateful node. 
We consider 
\emph{dependent inputs}, i.e., inputs that are themselves functions of other inputs to a stateful node. 
For each such candidate $i$, we query SKETCH to check if the function that computes $i$ can be \emph{folded into} the stateful node itself, such that the enlarged node fits into a stateful ALU. If the synthesis query is successful, $i$ is removed.
An example benchmark (BLUE (decrease)~\cite{blue}) where this works well is shown 
in Figure~\ref{fig:blue_decrease}. 
Here, \emph{folding} reduces an edge between the top two nodes,
thereby reducing the pipeline usage by 1. 

\emph{Predecessor packing to merge nodes.} Even after folding, the stateful update in a single node in $G$ might not fully utilize an available stateful ALU.  
Consider again the BLUE (decrease) example in Figure~\ref{fig:blue_decrease}, where the middle box shows $G$ after folding. 
Here, a single Tofino stateful ALU can actually implement \emph{both} stateful updates (in blue boxes) in a single stage,
as shown by a merged node on the right. 
To achieve this, we use a simple heuristic 
called \emph{predecessor packing}, inspired by technology mapping for hardware designs~\cite{flowmap}. The idea is to pack more into a stateful ALU by attempting a merge of nodes $u$ and $v$, where at least one node is stateful and where predecessor $u$ has only one out-edge (to $v$).
Again, we implement a merge-attempt via a SKETCH query, and merge the nodes if the query is successful. 
In our evaluations (\S\ref{ssec:eval_synthesis}), we show that these optimizations are effective in compiling to fewer stages in the pipeline.

\para{Step 3: Synthesizing stateful updates} 
To preserve the transactional semantics of the program, each stateful update must be completed within a single stage, 
i.e., it \emph{must} fit in a single stateful ALU.
For each stateful node in $G$, we generate a SKETCH query to check if the stateful update can be implemented by a stateful ALU using a grammar $A_1$. We assert that the query succeeds; if it fails, our procedure terminates with an error (and feedback to the user). 

\begin{figure} [t]
    \centering
    \includegraphics[width=1.05\linewidth]{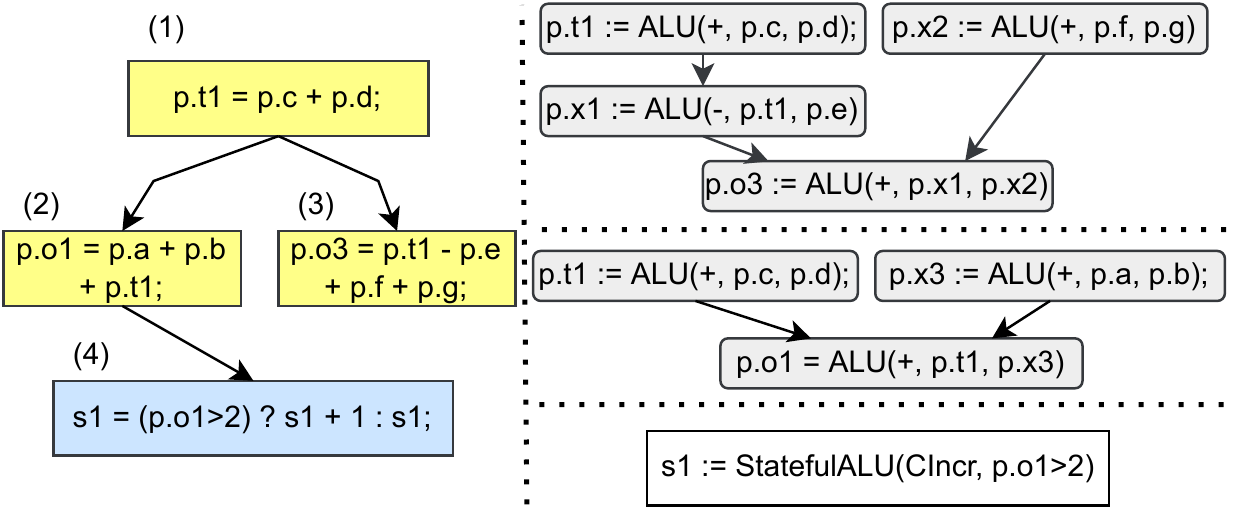}
    \caption{Example of a computation graph (on left) and the synthesis query results (on right) targeting Banzai ALUs. Stateful node is shown in blue and stateless nodes in yellow. The POs are:
    $\textsf{p.o1}$,$\textsf{p.o3}$. 
    The BCI of $\textsf{p.o1}$ contains nodes 1 and 2; the BCI of $\textsf{p.o3}$ contains nodes 1 and 3.  } 
    \label{fig:codegen_example}
\end{figure}

\para{Step 4: Minimum-depth solutions for stateless code} 
In the last step, we synthesize code for the inputs to the stateful ALUs and the POs of $G$. 
For each variable $o$ to be synthesized, we first compute 
its backwards cone of influence (BCI), 
which is often used in verification/synthesis tasks to determine the dependency region up to some (boundary of) inputs~\cite{SomenziBook}. 
Essentially, a BCI
provides the \emph{functional specification} for $o$ in terms of PIs and outputs of stateful nodes in $G$. Note that these specifications are stateless, i.e., they do not include any stateful nodes.
We model a switch's stateless ALU functionality using a grammar  $A_2$, and then use SKETCH to find a \emph{minimum-depth tree} solution for $o$, where each tree node represents a stateless ALU. Since SKETCH cannot directly optimize for depth, we invoke it in a loop.
An example computation graph with a single stateful update (blue box) and the associated synthesis query results are shown in Figure~\ref{fig:codegen_example}. 


\subsubsection{Final Result of Phase 2}
The final result of the synthesis procedure is represented in the form of a \emph{resource graph $R$} for a given P4 action block, where each node $v$ in $R$ represents a stateful/stateless ALU, and an edge $(u, v)$ in $R$ indicates that the output of ALU  $u$ is connected to an input of ALU $v$. These resource graphs play an important role in resource allocation, the next phase of our compiler. 
Our claim of correctness for the synthesis procedure along with a proof are provided in Appendix~\ref{appendix:correctnessThmProof}.

\subsection{Phase 3: Resource Allocation}
\label{ssec:resource_allocation}
After performing synthesis for each P4 action block, our third phase of the compiler performs \emph{global} resource allocation for the full P4 program
by using a constraint-based formulation,
shown in Table~\ref{table:phase3constraints}.
The top part lists the definitions of constants, indices, variables, and sets that are used to automatically generate the constraints.
The bottom part shows the full set of constraints, divided
into a first set that is similar to prior work~\cite{lavanya_nsdi15,cetus}, and a second set that is new in our work. 
Our new constraints address: (1) ALU resources in action computations, (2) multi-stage actions, (3) fitting multiple action blocks in the same pipeline stage, and (4) propagation of ALU outputs. 
Prior efforts either do not consider allocation of ALU resources and multi-stage actions~\cite{lavanya_nsdi15,cetus}, or do not address multiple action blocks~\cite{domino,chipmunk}. 
Another novel feature of our approach is that we use the resource graph $R$ synthesized for each action block (in Phase 2), 
to perform global optimization in this phase. 


\begin{table}[!t]
\scriptsize
\centering
\resizebox{\linewidth}{!}{  
\begin{tabular}{lll}
\toprule
  & \textbf{Name} & \textbf{Definition}\\
\midrule
\textbf{Constants} & $N_S$ & maximum number of pipeline stages \\  
& $N_{alu}$ & maximum number of ALUs in each stage \\
& $N_{H}$ & number of ALUs per stage for transporting packet header fields \\
& $N_{table}$ & maximum number of logical table IDs per stage \\
& $N_{entries}$ & maximum number of match entries per table per stage \\
& $e_t$ & maximum number of entries in table $t$ in program \\
\midrule
\textbf{Indices} & $t$ & index for Table \\
& $i$ & index for partition of a Table, partition denoted $t[i]$ \\ 
& $a$ & index for Action \\
& $u$ & index for ALU \\ 
& $s$ & index for pipeline Stage \\ 
\midrule
\textbf{Variables} & $M_{tis}$ & binary, set to 1 if match of $t[i]$  is assigned to stage $s$, 0 otherwise \\
& $stage_{u}$ & integer, stage assigned to ALU $u$ \\
& $stage_{us}$ & binary, set to 1 iff ALU $u$ is assigned to stage $s$ \\
& $beg_u$ & integer, stage where ALU $u$ output is computed \\
& $end_u$ & integer, $\geq$ last stage where ALU $u$ is used \\
& $prop_{us}$ & binary, set to 1 iff output of ALU $u$ is propagated in stage $s$ \\
\midrule
\textbf{Sets} & $R_{tia}$ & Resource graph for action $a$ of table partition $t[i]$ \\ 
 & $V_{tia}$ & Vertices in $R_{tia}$ \\ 
 & $E_{tia}$ & Edges in $R_{tia}$ \\ 
 & $Um$ & Set of ALUs whose output is propagated across stages \\
 & $Need_{u}$ & Set of ALUs that use ALU $u$ as input \\
\bottomrule
\end{tabular}}
\resizebox{\linewidth}{!}{  
\begin{tabular}{ll}
\toprule
\multicolumn{2}{l}{\textbf{Constraints similar to prior work}~\cite{lavanya_nsdi15,cetus}} \\ 
\midrule
Match table capacity &
$\forall s: \ \ \sum_{t, i} \ \ M_{tis} \leq N_{table}$ \\
Match action pairing & 
$\forall s \ \forall t, i, a \ \forall u \in V_{tia}: \ \ \ stage_{us} \rightarrow M_{tis}$ \\
Table dependency & 
$\forall i_1, i_2, a_1, a_2, \forall u_1 \in V_{t_1\ i_1\ a_1}, \ \forall u_2 \in V_{t_2\ i_2\ a_2}: stage_{u_1} < stage_{u_2}$ \\
\midrule
\multicolumn{2}{l}{\textbf{New Constraints in our work}} \\ 
\midrule
ALU allocation 1 & 
$ \forall t, i, a \ \forall u \in V_{tia}: \ \ 1 \leq stage_{u} \leq N_S$ \\
ALU allocation 2 & 
$\forall s\ \forall t, i, a,\ \forall u \in V_{tia}:\ \  stage_{u} = s \leftrightarrow stage_{us}$ \\
Action dependency &
$\forall t, i, a \ \forall (u, v) \in E_{tia}: \ \  stage_u < stage_v$ \\
ALU propagation 1 & 
$\forall u \in Um: \ \ beg_u = stage_u \land beg_u < end_u \land end_u \leq N_S$ \\
ALU propagation 2 & 
$\forall u \in Um, \forall v \in Need_u: \ \ end_u \geq stage_v$ \\
ALU propagation 3 & 
$\forall u \in Um, \forall s \in \{1,\ldots,N_S\}: \ \ (beg_u < s \land s < end_u) \leftrightarrow prop_{us}$ \\
ALU propagation 4 & 
$\forall s\ \forall u \in Um: \ \  stage_{us} + prop_{us} \leq N_{alu} - N_{H}$ \\
\bottomrule
\end{tabular}}
\caption{Constraint formulation for resource allocation.}
\label{table:phase3constraints}
\vspace{-0.25in}
\end{table}

\subsubsection{Constraints similar to prior work.}
If a match table in the program has too many entries to fit within a single stage, it is partitioned into $b_t$ separate tables, 
where $b_t = \lceil(e_t/N_{entries})\rceil$ (similar to prior work~\cite{lavanya_nsdi15,cetus}). 
If the match type is exact, a packet will match at most one of the partitions $t[i]$  
that have the same actions as table $t$. 
(In future work, we plan to support other types of matches such as ternary match.) 

The first constraint ensures that the number of match tables allocated in a stage is less than or equal to the number of table IDs available.
The second ensures that ALUs in action blocks are accompanied by the associated match table. 
The third enforces four types of table dependencies: match, action, successor, and reverse-match~\cite{lavanya_nsdi15}.
If table $t_2$ depends on table $t_1$, all ALUs of $t_2$ are allocated after ALUs of $t1$. 
For successor and reverse-match, $<$ is replaced by $\leq$.

\subsubsection{New constraints in our work.}
The constraints for ALU allocation (1,2) ensure that each ALU in each action is assigned to some pipeline stage~\footnote{By associating a binary and an integer variable to the assigned stage, we ensure its uniqueness without requiring additional constraints.}. 
The Action dependency constraint uses the edges in in $R_{tia}$ (synthesized in Phase 2) to enforce dependencies between ALUs. 
Together with the Table dependency constraint, this allows ALUs from multiple action blocks to be assigned in the same pipeline stage, \emph{while respecting both inter-table and intra-action dependencies}.
 
We support a multi-stage action under the condition that it 
does not modify the table's match key, by duplicating the match entries at each stage to ensure that the entire action is executed. As an example, suppose a match entry $m$ in table $t$ is associated with action $A$ that takes 2 stages.
We can allocate table $t$ in two consecutive stages, such that 
if a packet matches entry $m$ in table $t$ in stage $s$, it will match entry $m$ in table $t$ in stage $s+1$ as well, resulting in action $A$ being executed completely over the two stages. 

To allow flexibility in allocating multiple action blocks in the same pipeline stage, we also allow Action $A$ to be assigned to \emph{non-consecutive} stages. In this case, we need to allocate additional ALUs 
to \emph{propagate the intermediate results to other stages} (required by both Tofino and Menshen backends). 
The ALU propagation constraints (1-4) handle these additional ALUs, where the sets $U_m$, $Needs_u$ (defined in the top part of Table~\ref{table:phase3constraints}) are computed from $R_{tia}$.
(For ILP, ALU propagation 3 constraint can be formulated using the well-known Big-M method as shown in Appendix~\ref{appendix:bigm_encoding}.)

\subsubsection{Solving the constraint problem.}
We can use either an ILP solver (Gurobi~\cite{gurobi}) or an SMT solver (Z3~\cite{z3}, which supports optimization) to find an optimal or a feasible solution. 
We specify an objective function for finding an optimal solution, e.g., we add the constraint $\min\ cost$ to minimize the number of stages, where $cost$ is $\geq$ the stage assigned to any ALU.
i.e., $\forall t, i, a,\ \forall u \in V_{tia}:\ cost \geq stage_u$.
To find a feasible solution, we use a trivial objective function ($min\ 1$) with Gurobi (none is needed with Z3).


\section{Implementation and Evaluation}
\label{sec:eval}

We implement 
the CaT compiler with the workflow shown in Figure~\ref{fig:hls_workflow}. The resource transformation phase is implemented on top of p4c~\cite{p4c}. We also use p4c to identify the action blocks and table dependencies needed in CaT's resource synthesis and resource allocation phases. 
For the backend, ideally the CaT compiler should directly output machine code for the targets. However, due to the undocumented and proprietary machine code format of the Tofino chipset,
we generate a low-level P4 program by using a best-effort encoding for the 
resource constraints, based on known information about 
the Tofino chipset. 
For the Menshen backend, we extend the open-source RMT pipeline~\cite{menshen, menshen_site} by writing additional Verilog to support richer ALUs, 
e.g., the IfElseRAW ALU from the Domino paper~\cite{domino}. 
The CaT compiler directly outputs machine code to configure various programmable knobs (e.g., 
ALU opcodes, operands, etc.) within Menshen's Verilog code. 

\para{Sanity checking of CaT prototype} 
We check CaT's 
output for Menshen 
using its cycle-accurate simulator, which can be fed input packets to test the generated machine code. We create P4-16 benchmarks 
starting with a subset of the switch.p4 program~\cite{switch_p4}, 
consisting of 2--6 tables randomly sampled from switch.p4. 
Then, we add 
new actions to the tables 
using @atomic blocks for transactional behavior. 
The logic within these atomic blocks consists of one of 8 Domino benchmark programs~\cite{domino}. (The IfElseRaw ALU in our simulator is not expressive enough for the remaining 6.)
We also test the 8 benchmark programs in isolation,
generating 24 benchmarks in total, many of which have multiple transactions 
and thus stress both resource synthesis and resource allocation.
We randomly generate test input packets 
and 
inspect the output packets from the simulation. 
So far, all our sanity checks on CaT's outputs have passed. 

\subsection{Evaluation Setup and Experiments}
We address the following evaluation questions for components of the CaT compiler. 
    
    \textbf{Q1: Resource Transformation.} How much does the CaT's resource transformation help in terms of the resource usage? We select 3 
    benchmarks~\cite{transform_benchmarks} extracted from real P4 applications 
    and compare the resource usage for pre- and post-transformed programs
    (\S\ref{ssec:transform}).
    
    \textbf{Q2: Resource Synthesis.} How does CaT's resource synthesizer compare to existing ones? We compare CaT with Chipmunk over several dimensions (\S\ref{ssec:eval_synthesis}) using ALUs drawn from Tofino~\cite{p4sde} and Banzai~\cite{domino}.
    Appendix(\ref{ssec:codegen-exp}) shows a deeper analysis 
    on the benefits of the predecessor packing and preprocessing optimizations.
    
    \textbf{Q3: Resource Allocation.} How good is the CaT compiler in terms of resource usage? We use Gurobi as the default 
    for resource allocation 
    and compare the runtime of the Gurobi and Z3 solvers. In addition, 
    we compare the two modes of finding either an optimal or a feasible solution 
    (\S\ref{ssec:allocation-exp}).
    
    \textbf{Q4: Retargetable Backend.} Can CaT 
    easily perform 
    compilation for different hardware targets? 
    Our synthesis experiments with the Banzai and Tofino ALUs already demonstrate this feature. Additionally, we run the CaT compiler on different \emph{simulated} hardware configurations, compile the switch.p4 
    under 
    varying constraints and report the results (\S\ref{ssec:allocation-exp}).
\para{Benchmark selection}
\begin{itemize}
\vspace{0mm}
\item Resource transformation: 3 benchmarks (ME-1, ME-2, ME-3) extracted from SipHash and UPF (real P4 programs developed by P4 users). 
\item Resource synthesis: 14 benchmarks together with their semantically equivalent mutations (10 for each benchmark, so 140 in total) from the Chipmunk paper~\cite{chipmunk}.\footnote{Chipmunk can compile all 14 benchmarks by using Banzai ALUs~\cite{domino}, and 10 of the 14 benchmarks by using Tofino ALUs~\cite{chipmunk}. For Banzai ALUs, we also show the Domino pipeline usage as reported in the Chipmunk paper~\cite{chipmunk}.}
\item Resource allocation: Same as the benchmarks we use for sanity checking our prototype. We use the full switch.p4 program for experiments that vary hardware resource parameters in the Menshen backend.
\end{itemize}
\para{Machine configuration} 
We use a 4-socket AMD Opteron 6272 (2.1 GHz) with 64 hyperthreads and 256 GB RAM to run all our experiments for both CaT and Chipmunk. Note that Chipmunk runs more complex SKETCH queries and
 uses many identical machines to concurrently explore different-sized ALU grids,
whereas CaT only uses one machine, 
saving a lot of computation resources.

\begin{table*}[!t]
\scriptsize
\centering
\resizebox{\textwidth}{!}{  
\begin{tabular}{lllllllllllc}
\toprule
\multicolumn{1}{c|} {\textbf{Program}} & \multicolumn{1}{c|} {\textbf{ALU}} & \multicolumn{5}{c|}{\textbf{CaT}} & \multicolumn{3}{c|}{\textbf{Chipmunk \cite{chipmunk}}} &\multicolumn{1}{c|} {\textbf{CaT Speedup}} &{\textbf{Domino \cite{domino}}}\\ 
\multicolumn{1}{c|} {} & \multicolumn{1}{c|} {} & Avg Time (s) & Std Time (s) & \multicolumn{3}{c|} {Avg \#stages} & Avg Time (s) & Std Time (s) & \multicolumn{1}{c|} {Avg \#stages} & \multicolumn{1}{c|}{wrt Chipmunk} & Avg \#stages \\
\multicolumn{1}{c|} {} & \multicolumn{1}{c|} {} & & & \textbf{default} & w/o pred & \multicolumn{1}{c|} {w/o ppa} & & & \multicolumn{1}{c|} {} & \multicolumn{1}{c|} {} & (from \cite{chipmunk})\\
\midrule
BLUE (increase) \cite{blue} & \textbf{Tofino ALU} & 19.04 & 0.43 & 1 & \cellcolor{FairC}2 & 1 & 159.78 & 59.03 & 2 & \cellcolor{GoodC} 8.39 $\times$ & \cellcolor{BadC}\\
BLUE (decrease) \cite{blue} & \textbf{Tofino ALU} & 18.72 & 0.84 & 1 & \cellcolor{FairC}2 & 1 & 142.5 & 42.5 & 2 & \cellcolor{GoodC} 7.61 $\times$ & \cellcolor{BadC}\\
Flowlet switching \cite{flowlet} & \textbf{Tofino ALU} & 19.76 & 0.69 & 2 & \cellcolor{FairC}\xmark & 2 & 962.83 & 1170.16 & 2 & \cellcolor{GoodC} 48.73 $\times$ & \cellcolor{BadC}\\
Marple new flow \cite{marple} & \textbf{Tofino ALU} & 6.65 & 0.52 & 1 & \cellcolor{FairC}\xmark & 1 & 5.2 & 1.71 & 1 & \cellcolor{BadC} 0.78 $\times$ & \cellcolor{BadC}\\
Marple TCP NMO \cite{marple} & \textbf{Tofino ALU} & 13.24 & 0.53 & 2 & \cellcolor{FairC}\xmark & \cellcolor{FairC}\xmark & 6.56 & 0.36 & 2 & \cellcolor{BadC} 0.50 $\times$ & \cellcolor{BadC} \textbf{N/A}\\
Sampling \cite{domino} & \textbf{Tofino ALU} & 14.03 & 0.57 & 1 & \cellcolor{FairC}\xmark & 1 & 22.87 & 10.68 & 1 & \cellcolor{GoodC} 1.63 $\times$ & \cellcolor{BadC}\\
RCP \cite{rcp} & \textbf{Tofino ALU} & 20.19 & 0.59 & 1 & \cellcolor{FairC}\xmark & 1 & 65.13 & 20.93 & 1 & \cellcolor{GoodC} 3.23 $\times$ & \cellcolor{BadC}\\
SNAP heavy hitter \cite{snap} & \textbf{Tofino ALU} & 3.58 & 0.25 & 1 & 1 & 1 & 26.83 & 13.63 & 1 & \cellcolor{GoodC} 7.49 $\times$ & \cellcolor{BadC}\\
DNS TTL change \cite{dns_ttl} & \textbf{Tofino ALU} & 20.84 & 1.97 & 2 & \cellcolor{FairC}3 & \cellcolor{FairC}\xmark & 36.34 & 50.55 & 2 & \cellcolor{GoodC} 1.74 $\times$ & \cellcolor{BadC}\\
CONGA \cite{conga} & \textbf{Tofino ALU} & 10.24 & 0.43 & 1 & \cellcolor{FairC}\xmark & 1 & 3.02 & 0.17 & 1 & \cellcolor{BadC} 0.29 $\times$ & \cellcolor{BadC}\\
\midrule
BLUE (increase) \cite{blue} & \textbf{Banzai ALU:} pred raw & 40.69 & 1.41 & 4 & 4 & 4 & 166.88 & 36.59 & 4 & \cellcolor{GoodC} 4.10 $\times$ & \cellcolor{BadC}\xmark\\
BLUE (decrease) \cite{blue} & \textbf{Banzai ALU:} sub & 38.83 & 1.48 & 4 & 4 & 4 & 1934.82 & 1611.66 & 4 & \cellcolor{GoodC} 49.83 $\times$ & \cellcolor{BadC}\xmark\\
Flowlet switching \cite{flowlet} & \textbf{Banzai ALU:} pred raw & 25.37 & 0.94 & 3 & 3 & 3 & 185.84 & 81.41 & 3 & \cellcolor{GoodC} 7.33 $\times$ & \cellcolor{BadC}8.3\\
Marple new flow \cite{marple} & \textbf{Banzai ALU:} pred raw & 13.79 & 0.44 & 2 & 2 & 2 & 12.31 & 0.18 & 2 & \cellcolor{BadC} 0.89 $\times$ & \cellcolor{BadC}\xmark\\
Marple TCP NMO \cite{marple} & \textbf{Banzai ALU:} pred raw & 28.12 & 2.60 & 3 & \cellcolor{FairC}4 & \cellcolor{FairC}\xmark & 15.3 & 0.49 & 3 & \cellcolor{BadC} 0.54 $\times$ & \cellcolor{BadC}\xmark\\
Sampling \cite{domino} & \textbf{Banzai ALU:} if else & 11.52 & 0.65 & 2 & 2 & 2 & 33.39 & 11.09 & 2 & \cellcolor{GoodC} 2.90 $\times$ & \cellcolor{BadC}\xmark\\
RCP \cite{rcp} & \textbf{Banzai ALU:} pred raw & 25.08 & 0.85 & 2 & 2 & 2 & 31.21 & 7.55 & 2 & \cellcolor{GoodC} 1.24 $\times$ & \cellcolor{BadC}5.6\\
SNAP heavy hitter \cite{snap} & \textbf{Banzai ALU:} pair & 3.45 & 0.23 & 1 & 1 & 1 & 69.07 & 19.36 & 1 & \cellcolor{GoodC} 20.02 $\times$ & \cellcolor{BadC}3.3\\
DNS TTL change \cite{dns_ttl} & \textbf{Banzai ALU:} nested if & 32.63 & 34.91 & 3
& \cellcolor{FairC}5 & \cellcolor{FairC}\xmark & 211.67 & 22.65 & 3 & \cellcolor{GoodC} 6.49 $\times$ & \cellcolor{BadC}\xmark\\
CONGA \cite{conga} & \textbf{Banzai ALU:} pair & 10.27 & 0.55 & 1 & 1 & 1 & 19.47 & 8.05 & 1 & \cellcolor{GoodC} 1.90 $\times$ & \cellcolor{BadC}\xmark\\
Stateful firewall~\cite{snap} & \textbf{Banzai ALU:} pred raw & 2499.43 & 4638.58 & 4 & 4 & \cellcolor{FairC}\xmark & 6749.89 & 6349.94 & 4 & \cellcolor{GoodC} 2.70 $\times$ & \cellcolor{BadC}15.5\\
Learn filter~\cite{domino} & \textbf{Banzai ALU:} raw & 31.01 & 0.73 & 3 & 3 & 3 & 212.32 & 4.47 & 3 & \cellcolor{GoodC} 6.85 $\times$ & \cellcolor{BadC}17.5\\
Spam Detection~\cite{snap} & \textbf{Banzai ALU:} pair & 3.51 & 0.21 & 1 & 1 & 1 & 59.95 & 17.75 & 1 & \cellcolor{GoodC} 17.08 $\times$ & \cellcolor{BadC}3.1\\
STFQ~\cite{stfq} & \textbf{Banzai ALU:} nested if & 20.99 & 2.04 & 3 & 3 & 3 & 22.73 & 6.94 & 2 & \cellcolor{GoodC} 1.08 $\times$ & \cellcolor{BadC}\xmark\\
\bottomrule
\end{tabular}}
\begin{small}
\caption{CaT vs. Chipmunk and Domino; with Tofino or Banzai ALUs. (pred: Predecessor packing, ppa: Preprocessing, \xmark: failed)}
\label{table:Code_generation_time_comparison_domino_tofino}
\end{small}
\end{table*}

\subsection{Results for Resource Transformation}
\label{ssec:transform}
The resource transformation phase of CaT compiler rewrites if-else statements in the P4 program into match-action tables as much as possible. For benchmarks ME-1, ME-2, and ME-3, the post-transformed programs reduced the usage of gateway resource. In addition, since the rewriting merges many default tables 
from if-else branches
into one big table, the total number of tables is reduced. 
We also confirm that the transformed programs use fewer stages than the original programs by compiling them using the Tofino compiler.


\subsection{Results for Resource Synthesis}
\label{ssec:eval_synthesis}

We compare the CaT and Chipmunk compilers on the SipHash benchmark (cf. Figure~\ref{fig:motivation_siphash}) and on all benchmarks used in the Chipmunk work~\cite{chipmunk}. 
For the latter, we consider 
targeting both Tofino ALUs and Banzai ALUs, 
to test the performance of CaT on different instruction sets  and 
various input programs. This also demonstrates CaT's retargetability via different ALU grammars.

\para{Results for SipHash} 
CaT successfully compiles the SipHash P4 program~\cite{siphash} with Tofino ALU in around 40 hours. This runtime is much higher than for other benchmark examples due to two main reasons. First, the compiled program takes 4 pipeline stages due to a single multistage action, resulting in a large synthesis search space. Second, the SipHash example involves bitvector operations in addition to integer arithmetic, making the synthesis problem harder for SKETCH (the program uses 32-bit bitvectors, whereas SKETCH's default for integers is 5 bits). 
To reduce the CaT compile time, we experiment with further decomposing the synthesis queries 
in the computation graph, after which the total time for synthesis is reduced to $\approx$10s. 
In comparison, Chipmunk fails to output the result even after 150 hours.

\para{Results for Chipmunk benchmarks} 
The results are shown in Table~\ref{table:Code_generation_time_comparison_domino_tofino}, for Tofino ALUs and Banzai ALUs, respectively. We report the runtime of full compilation; for CaT, this includes the resource allocation time, which Chipmunk does not perform. We consider 10 semantically equivalent mutations of each of the benchmarks, and report the mean and standard deviation of compilation time across all mutations. \footnote{
The running times 
in Table~\ref{table:Code_generation_time_comparison_domino_tofino} are similar to, but slightly different from that in Table 2 of the Chipmunk paper~\cite{chipmunk}. The differences arise due to Chipmunk's use of SKETCH's parallel mode, which is
non-determinism.} 

We also report the resource consumption of the generated code produced by CaT, in terms of the total number of pipeline stages required on average across mutations ("\#stages default" column).
This is the most important scarce resource in network substrates 
(e.g., 12 stages for Tofino). For evaluating the effectiveness of our predecessor packing optimization (pred) and the preprocessing analyses (ppa) (\S\ref{ssec:codegen}), we also report the number of stages without these turned on in columns "\#stages w/o pred", "\#stages w/o ppa", respectively. Gray-ed entries indicate a difference from the default setting (more detailed analysis in Appendix~\ref{ssec:codegen-exp}).

Our results show that CaT is able to compile \emph{all} programs
successfully compiled by Chipmunk, with almost all compiled results having a matching number of pipeline stages. Furthermore,  \emph{CaT is often much faster and more stable (in running time) than Chipmunk}. Specifically, for the Tofino ALUs (Table~\ref{table:Code_generation_time_comparison_domino_tofino}), CaT finishes 
compilation 
within a few seconds, 
$2.75$x faster on average (geometric mean) than Chipmunk. The max speedup is 48x for \emph{flowlet switching}, 
a minutes-to-seconds improvement ($\approx$16 minutes in Chipmunk vs. 20 seconds in CaT). 
In \emph{BLUE(increase)} and \emph{BLUE(decrease)}, CaT generates a solution with \emph{less} number of stages than Chipmunk. In all other benchmarks the number of stages is the same. 
For the BLUE benchmarks, 
since the Tofino stateful ALU contains two registers, CaT's optimizations enabled it to pack a successive pair of stateful updates 
into a single stateful ALU (\S\ref{ssec:codegen}). 
In comparison, Chipmunk mapped the two stateful updates to two ALUs in two stages. This shows that CaT's approach can find additional opportunities for fully utilizing the functionality of available hardware resources. Predecessor packing is also effective in 9 of 10 benchmarks, enabling compilation to succeed or reducing the number of stages; preprocessing is also useful in 2 benchmarks.

For the Banzai ALUs (Table~\ref{table:Code_generation_time_comparison_domino_tofino}),
we additionally report results on stages required by the Domino compiler~\cite{domino} (that only handled Banzai ALUs), with the average number of stages across different program mutations shown in the last column (as reported in~\cite{chipmunk}).
Note first that CaT takes no more than 1 minute on most successful benchmarks. Although it takes 40 minutes 
for \emph{stateful firewall}, Chipmunk is much slower,  requiring more than 1.5 hours. 
CaT provides $3.94$x speedup on average (geometric mean) and $49$x maximum, with respect to Chipmunk. 
CaT is slower only on \emph{Marple new flow} and \emph{Marple TCP NMO}, but finishes both within 30 seconds. Note that Chipmunk uses multiple machines and only reports the minimum running time across all, while CaT only uses one machine to do the synthesis.
In terms of number of stages, CaT generates code with the same number of stages as Chipmunk for all benchmarks except the \emph{STFQ} example (3 in CaT vs. 2 in Chipmunk).
Upon investigation, we find that this is due to separation between queries for stateful and stateless nodes in our synthesis procedure. Although our predecessor packing optimization can often mitigate this negative effect, we plan to improve it further in future work. 
Still, both predecessor packing and preprocessing optimizations are effective in some benchmarks here as well.
Finally, note that Domino either fails to compile (8 of 14 examples), or uses many more stages (other 6 examples).
\emph{Overall, 
our results show that CaT 
can generate high-quality code comparable to Chipmunk, but in much less time and with less powerful compute resources.}
\subsection{Results for Resource Allocation}
\label{ssec:allocation-exp}
We 
experiment with two solvers (Gurobi and Z3) and two modes (optimal and feasible
) on our benchmark examples. The detailed data (CDFs of runtime and 
stage count) are shown in Appendix~\ref{appendix:resource_allocation}  
(Figure~\ref{fig:gurobiVsZ3}). 
The results show that for checking feasibility, Gurobi returns suboptimal solutions that use all the pipeline stages, while Z3 finds feasible solutions that are better than Gurobi's but takes marginally more time. 
However, Gurobi finds an optimal solution in almost the same time 
as it takes to find a feasible solution. For these benchmarks, Gurobi is 
faster than Z3. Thus, Gurobi with optimization is a good default. 

We also study the resource allocation time of switch.p4 as a function of the parameters of the 
Menshen backend target. 
We vary the maximum number of entries per table, number of stages, and number of tables per stage,
and plot the runtime of Gurobi in both optimal and feasible mode. 
The results (in Appendix~\ref{appendix:resource_allocation}, 
Figures~\ref{fig:entries},~\ref{fig:stage},~\ref{fig:tables}) 
show that the runtime is similar for optimal and feasible modes, but varies significantly on different instances depending on whether there is a solution. 

\section{Conclusions}
\label{sec:conclusion}
This paper introduces a new approach to building compilers for packet-processing pipelines based on high-level synthesis. 
We adopt this approach to develop \sysname, a compiler for P4 programs. 
\sysname can handle more programs, reduce pipeline resource usage, and compile faster and with fewer compute resources than existing compilers.
We hope these results encourage compiler engineers for such packet-processing pipelines to adopt similar ideas for production-quality compilers. 

{\bf This work does not raise any ethical issues.}


\balance
\bibliographystyle{plain}
\bibliography{reference}
\appendix

\section{Pseudo-code for Synthesis Procedure}
\label{appendix:alg1}

Our synthesis procedure is shown in Alg.~1. 
It consists of four main steps: 1) normalization; 2) folding and predecessor packing optimizations; 3) synthesis of stateful updates; 4) synthesis of minimum-depth solutions for stateless code. 

\begin{algorithm*} [h]
    \centering
    \vspace*{-0.05in}
    \includegraphics[width=0.95\linewidth]{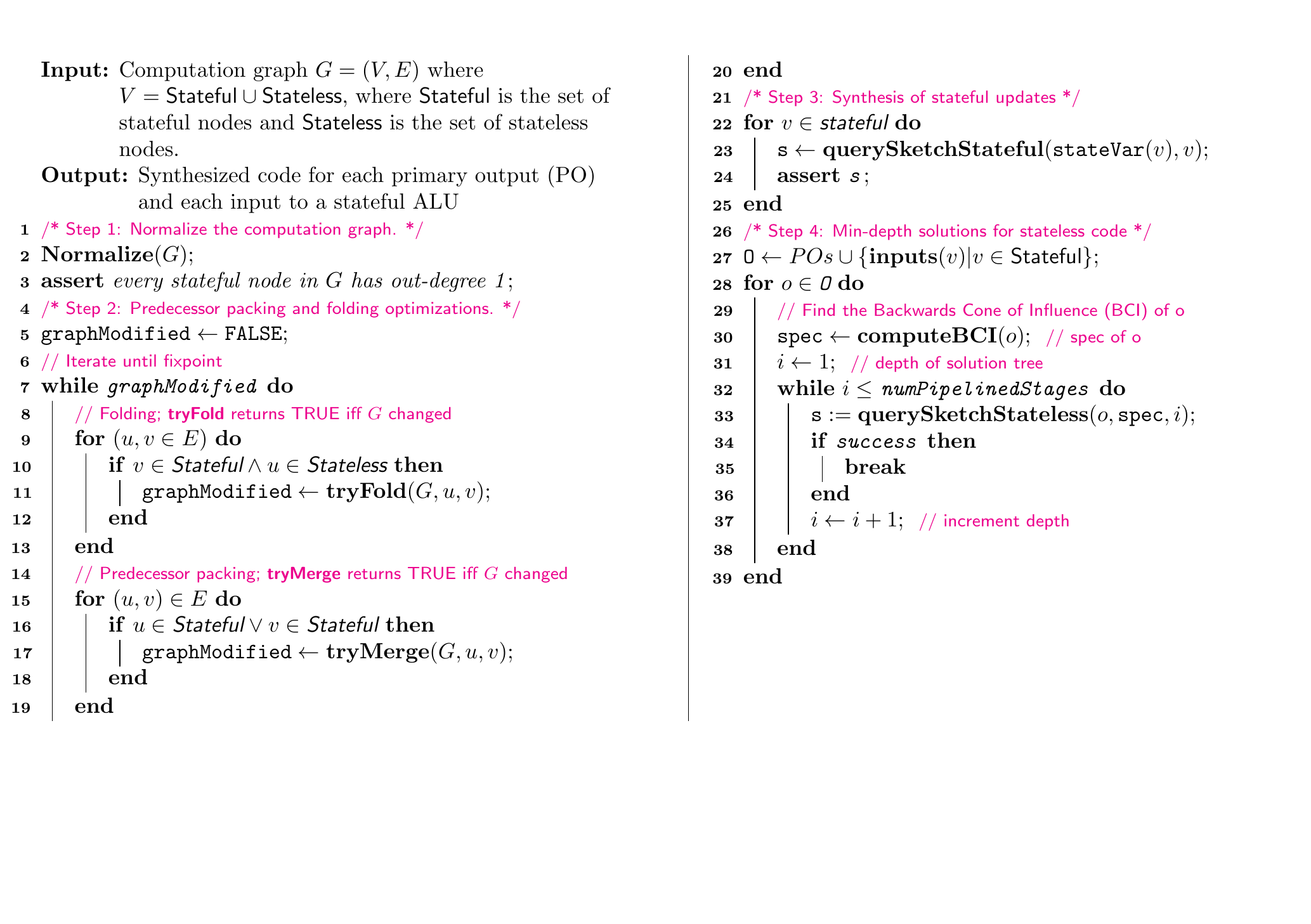}
    \vspace*{-1in}
   \caption{Code Generation Algorithm for CaT.}
    \label{alg:1}
    \vspace*{0.2in}
\end{algorithm*}

\section{Correctness of Synthesis Procedure}
\label{appendix:correctnessThmProof}

\newtheorem{thm1}{Theorem}
\begin{thm1}[Correctness]
\label{thm:correct}
The result of the CaT synthesis procedure (Alg.~1) on a computation graph $G$ is correct.
\end{thm1}

\begin{proof}[Proof Sketch]
Our synthesis procedure works by decomposing $G$ (after correctness-preserving normalization and optimizations in Steps 1 and 2 of Alg.~1, respectively) into disjoint subgraph components of stateful nodes and stateless BCIs of POs. Each such subgraph of $G$ corresponds to a \emph{specification} of either a PO or a stateful upate. Each of these is then synthesized into a subgraph of $R$, by Steps 3 and 4 of Alg.~1, respectively. Each synthesized subgraph corresponds to an \emph{implementation} of the specification of either a PO or a stateful update. Based on correctness of program synthesis in SKETCH~\cite{sketch}, the stateful updates and POs in $R$ are functionally equivalent to those in $G$. Hence the resource synthesis result is correct.
\end{proof}

\section{ILP encoding for ALU propagation constraints}
\label{appendix:bigm_encoding}
We use the big-M method to obtain an ILP formulation of the constraint
\begin{equation*}
    \forall u \in I, \forall s \ (beg_u < s \land s < end_u) \leftrightarrow prop_{us} = 1
\end{equation*}
For each $u \in I$ and $s \in \{1, \ldots N_S\}$, we use a binary variable $lo_{us}$ as an indicator for $beg_u < s$ and a binary variable $hi_{us}$ as an indicator for $s < end_u$. $M$ is a large constant (e.g., $N_S + 5$).

The following constraints ensure that $lo_{us}$ is $1$ if $beg_u < s$ and $0$ otherwise.
\begin{align}
    s - beg_u \leq M lo_{us} \label{eq:lo1}\\
    s - beg_u > -M(1 - lo_{us}) \label{eq:lo2}
\end{align}
If $s - beg_u > 0$ then $lo_{us} = 1$~\eqref{eq:lo1} and if $s - beg_u \leq 0$ then $lo_{us} = 0$~\eqref{eq:lo2}.

The following constraints ensure that $hi_{us}$ is $1$ if $s < end_u$ and $0$ otherwise.
\begin{align}
    s - end_u < M(1 - hi_{us}) \label{eq:hi1}\\
    s - end_u \geq -M hi_{us} \label{eq:hi2}
\end{align}
If $s - end_u < 0$ then $hi_{us} = 1$~\eqref{eq:hi2} and if $s - end_u \geq 0$ then $hi_{us} = 0$~\eqref{eq:hi1}.

The following constraints use $lo_{us}$ and $hi_{us}$ to make $prop_{us}$ an indicator for $beg_u < s < end_u$.
\begin{align}
    lo_{us} + hi_{us} - 2 < M prop_{us} \label{eq:prop1} \\
    lo_{us} + hi_{us} - 2 \geq -M (1 - prop_{us}) \label{eq:prop2}
\end{align}
If $lo_{us} + hi_{us} - 2 \geq 0$ then $prop_{us} = 1$~\eqref{eq:prop1} and if $lo_{us} + hi_{us} - 2 < 0$ then $prop_{us} = 0$~\eqref{eq:prop2}. This means that $prop_{us} = 1$ only if both $lo_{us} = 1$ and $hi_{us} = 1$. Hence, $prop_{us} = 1$ if $s > beg_u$ and $s < end_u$, otherwise $prop_{us} = 0$.

\section{Discussion of resource synthesis results}
\label{ssec:codegen-exp}

We performed controlled experiments to evaluate two optimizations (\S\ref{ssec:codegen}): (1) Predecessor packing, and (2) Preprocessing analyses (constant folding, algebraic simplification, dead code elimination). The results for the Banzai ALUs are shown in Table~\ref{table:Code_generation_time_comparison_domino_tofino}
for CaT (with both optimizations enabled), CaT without predecessor packing (CaT w/o pred), and CaT without preprocessing analyses (CaT w/o ppa). 

\paragraph{Predecessor packing.} 
As for Banzai ALUs, without predecessor packing, CaT uses additional stages to compile two examples (\emph{Marple TCP NMO}, and \emph{DNS TTL change}). 
, showing that predecessor packing can reduce the number of pipeline stages.
With the Tofino ALU (results in Table~\ref{table:Code_generation_time_comparison_domino_tofino}
), predecessor packing was even more beneficial: disabling predecessor packing resulted in compilation errors for 6 examples (\emph{flowlets, Marple new flow, Marple TCP NMO, Sampling, RCP}, and \emph{CONGA}), since the Tofino ALU supports very limited stateless computation and cannot handle relational or conditional expressions. Packing such expressions into adjacent stateful ALUs allowed compilation to succeed.
\paragraph{Preprocessing analyses.}
As for Banzai ALUs, without preprocessing analyses, three of the examples could not be compiled. 
For all other examples, the number of stages is identical with and without this optimization. We observe similar trends when the Tofino ALU is used (Table~\ref{table:Code_generation_time_comparison_domino_tofino}). 
The runtime of the preprocessing analyses itself is negligible, less than 0.1 sec in all examples.
\emph{Overall, our results show the effectiveness of these optimizations in providing a sweet spot for CaT, such that it is much faster than Chipmunk while generating optimal or close-to-optimal number of stages.} 

\section{Additional results for resource allocation}
\label{appendix:resource_allocation}

\begin{table*}[h]
\scriptsize
\centering
\begin{tabular}{lllllllll}
\toprule
\textbf{Benchmark} & \multicolumn{2}{c}{\textbf{Gurobi opt}} & \multicolumn{2}{c}{\textbf{Gurobi sat}} & \multicolumn{2}{c}{\textbf{Z3 opt}} & \multicolumn{2}{c}{\textbf{Z3 sat}}\\ 
 & Time (s) & Stages & Time (s) & Stages & Time (s) & Stages & Time (s) & Stages \\
 \midrule
stateful fw & 0.133 & 4 & 0.14 & 12 & 0.251 & 4 & 0.273 & 4\\
Blue increase & 0.1 & 4 & 0.12 & 12 & 0.218 & 4 & 0.244 & 4\\
marple new flow & 0.102 & 2 & 0.11 & 12 & 0.201 & 2 & 0.225 & 2\\
sampling & 0.102 & 2 & 0.103 & 12 & 0.21 & 2 & 0.226 & 2\\
flowlets & 0.127 & 3 & 0.124 & 12 & 0.232 & 3 & 0.254 & 3\\
rcp & 0.131 & 2 & 0.135 & 12 & 0.234 & 2 & 0.253 & 2\\
learn\_filter & 0.151 & 3 & 0.146 & 12 & 0.241 & 3 & 0.272 & 3\\
marple\_tcp & 0.117 & 3 & 0.119& 12 & 0.223 & 3 & 0.241 & 3\\
benchmark9 & 0.169 & 4 & 0.161 & 12 & 0.81 & 4 & 1.8 & 12\\
benchmark10 & 0.165 & 4 & 0.163 & 12 & 0.781 & 4 & 0.501 & 4\\
benchmark11 & 0.19 & 4 & 0.18 & 12 & 1.13 & 4 & 0.639 & 12\\
benchmark12 & 0.178 & 4 & 0.168 & 12 & 0.967 & 4 & 0.502 & 11\\
benchmark13 & 0.18 & 3 & 0.173 & 12 & 0.88 & 3 & 0.506 & 11\\
benchmark14 & 0.202 & 3 & 0.182 & 12 & 0.889 & 4 & 0.523 & 11\\
benchmark15 & 0.241 & 4 & 0.22 & 12 & 1.033 & 4 & 0.585 & 11\\
benchmark16 & 0.198 & 4 & 0.183 & 12 & 0.963 & 4 & 0.553 & 12\\
benchmark17 & 0.159 & 3 & 0.142 & 12 & 0.845 & 4 & 0.459 & 12\\
benchmark18 & 0.156 & 3 & 0.146 & 12 & 0.922 & 4 & 0.447 & 12\\
benchmark19 & 0.169 & 4 & 0.154 & 12 & 0.946 & 4 & 0.531 & 12\\
benchmark20 & 0.181 & 5 & 0.17 & 12 & 1.086 & 4 & 0.6 & 11\\
benchmark21 & 0.145 & 3 & 0.148 & 12 & 0.733 & 4 & 0.462 & 12\\
benchmark22 & 0.162 & 3 & 0.168 & 12 & 1.578 & 4 & 1.578 & 12\\
benchmark23 & 0.215 & 3 & 0.201 & 12 & 2.544 & 4 & 0.798 & 12\\
benchmark24 & 0.174 & 3 & 0.149 & 12 & 1.53 & 4 & 0.595 & 12\\
\bottomrule
\end{tabular}
\begin{small}
\caption{Comparing optimal and feasible for Gurobi and Z3\label{table:ilp}}
\end{small}
\vspace{-0.1in}
\end{table*}

We experiment with two solvers (Gurobi vs. Z3) and two modes (optimal and feasible solutions) on all our 24 benchmarks. We report both time spent running the solvers and the final number of stage usage to compare between different solvers and different modes. Table~\ref{table:ilp} shows the detailed results.

\begin{figure*}[!t]
\begin{minipage}[b]{0.33\linewidth}
    \centering
    \includegraphics[width=\linewidth]{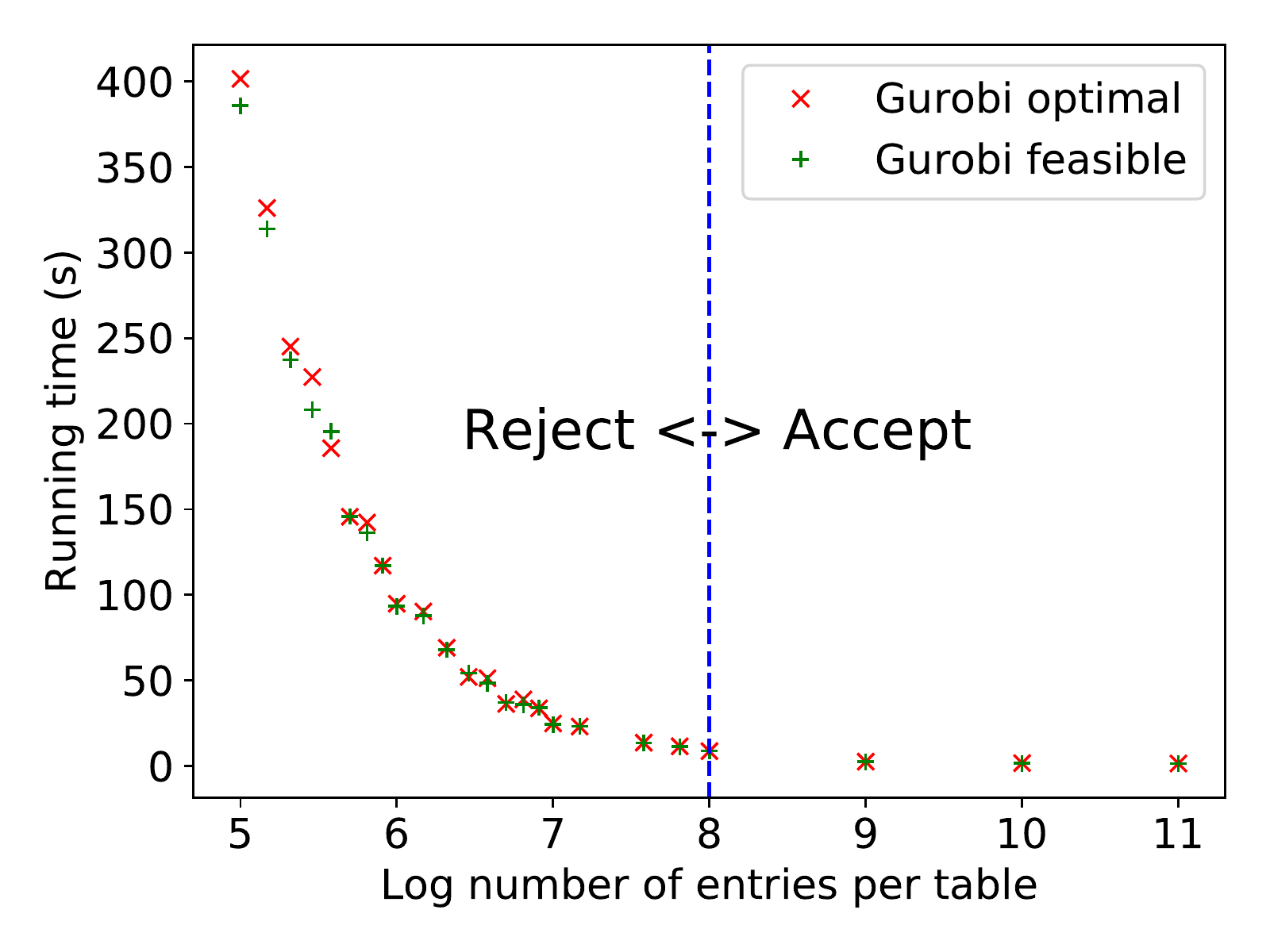}
    \begin{tiny}
    \caption{Varying \# of max. entries/table.\label{fig:entries}}
    \end{tiny}
\end{minipage}\hfill
\begin{minipage}[b]{0.33\linewidth}
    \centering
    \includegraphics[width=\linewidth]{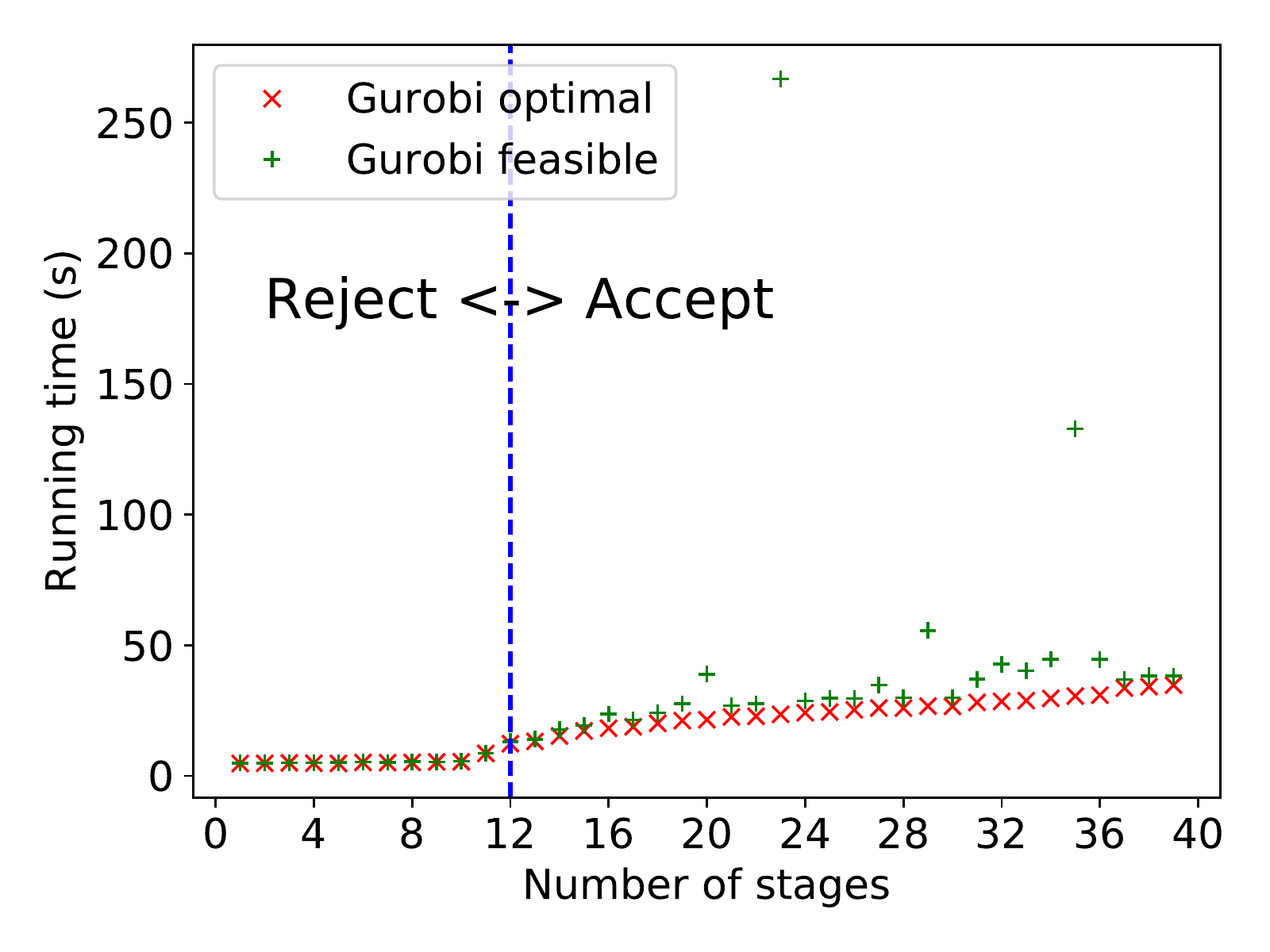}
    \begin{tiny}
    \caption{Varying \# of stages.\label{fig:stage}}
    \end{tiny}
\end{minipage}
\begin{minipage}[b]{0.33\linewidth}
    \centering
    \includegraphics[width=\linewidth]{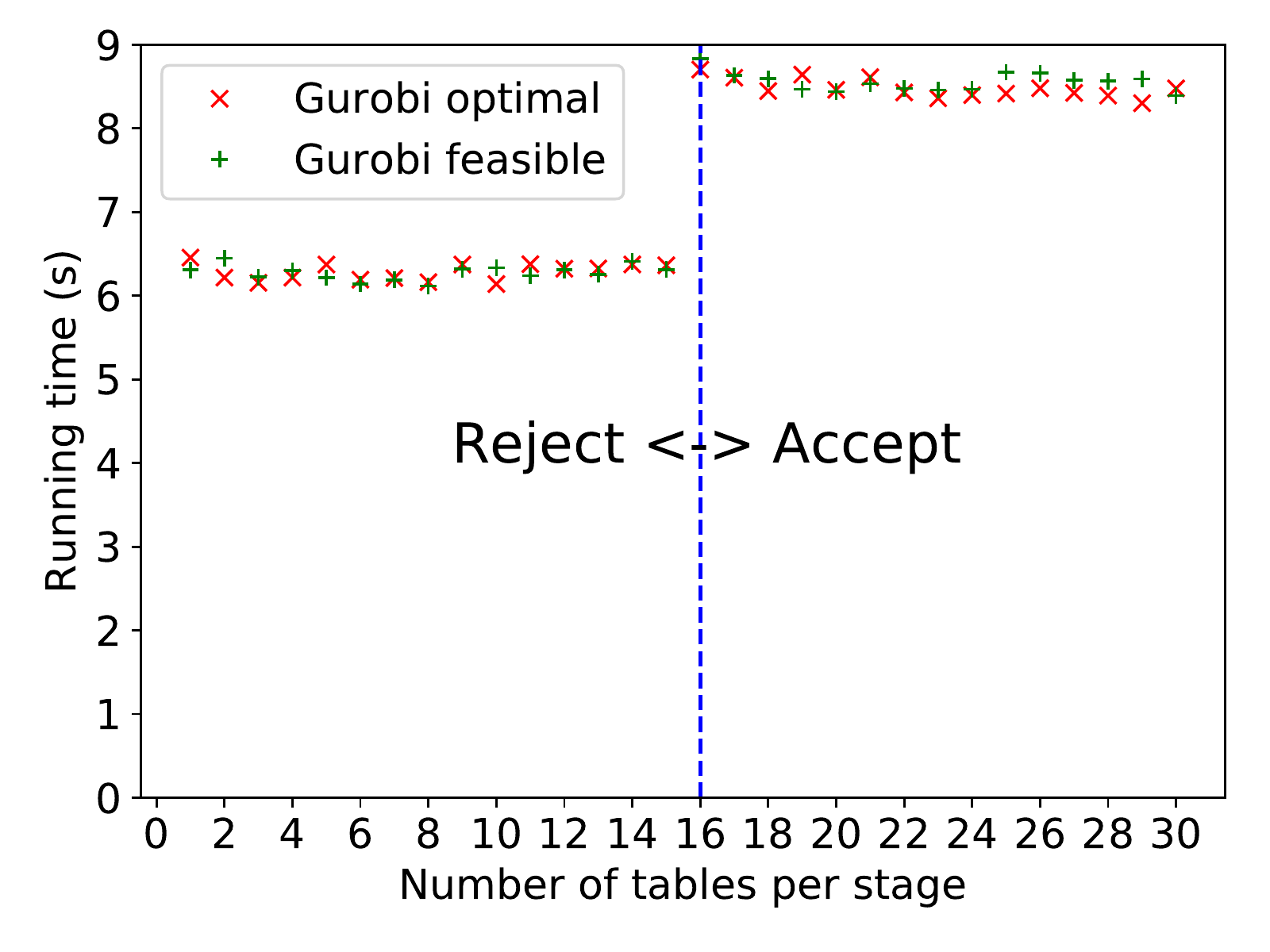}
    \begin{tiny}
    \caption{Varying \# of tables per stage.\label{fig:tables}}
    \end{tiny}
\end{minipage}
\vspace*{-0.1in}
\label{fig:resource-alloc-hwparams}
\end{figure*}

We also run experiments on different simulated hardware configurations for finding feasible vs. optimal solutions using Gurobi. The results are shown in Figures~\ref{fig:entries}, ~\ref{fig:stage}, ~\ref{fig:tables}. We plot a vertical line indicating the transition from reject to accept for the constraint solver. Across a variety of hardware configurations, we find that the runtime of both modes are quite similar. Figure \ref{fig:entries} shows that the runtime increases as the maximum number of entries decreases because of the increase in the number of partitions of a table (and hence number of variables) as the maximum number of entries decreases. Figure \ref{fig:stage} shows that runtime as the number of stages increases because of the increase in the number of indicator variables tracking which stage a table belongs to. In Figure~\ref{fig:tables}, the number of Gurobi variables is constant as we vary the number of tables per stage; however, whether a program can be rejected or accepted given the hardware constraints affects the runtime of the ILP procedure.

\begin{figure*}[!t]
\begin{minipage}[b]{0.48\linewidth}
    \centering
    \includegraphics[width=\linewidth]{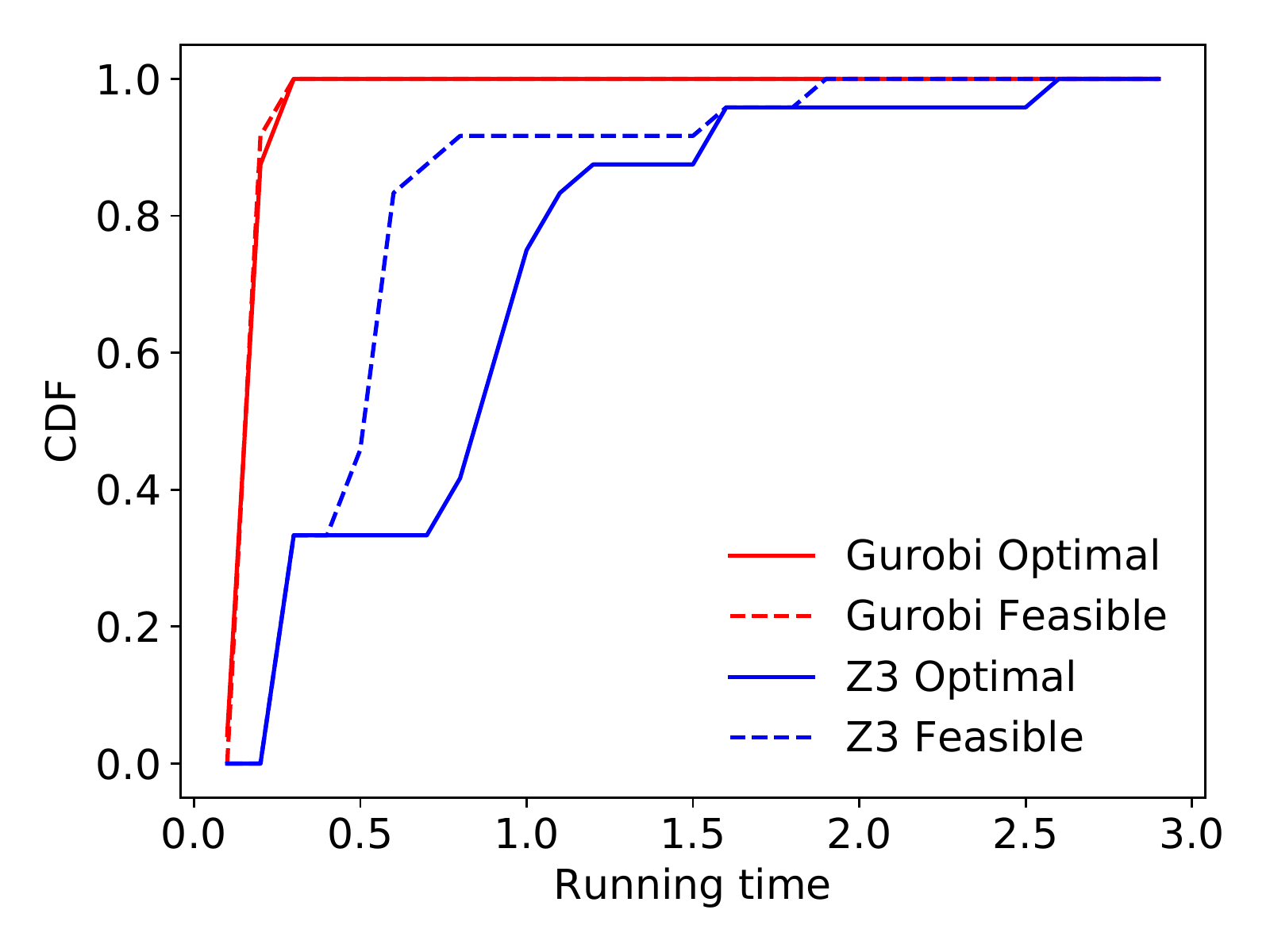}
    \begin{tiny}
    \label{fig:run_time}
    \end{tiny}
\end{minipage}\hfill
\begin{minipage}[b]{0.48\linewidth}
    \centering
    \includegraphics[width=\linewidth]{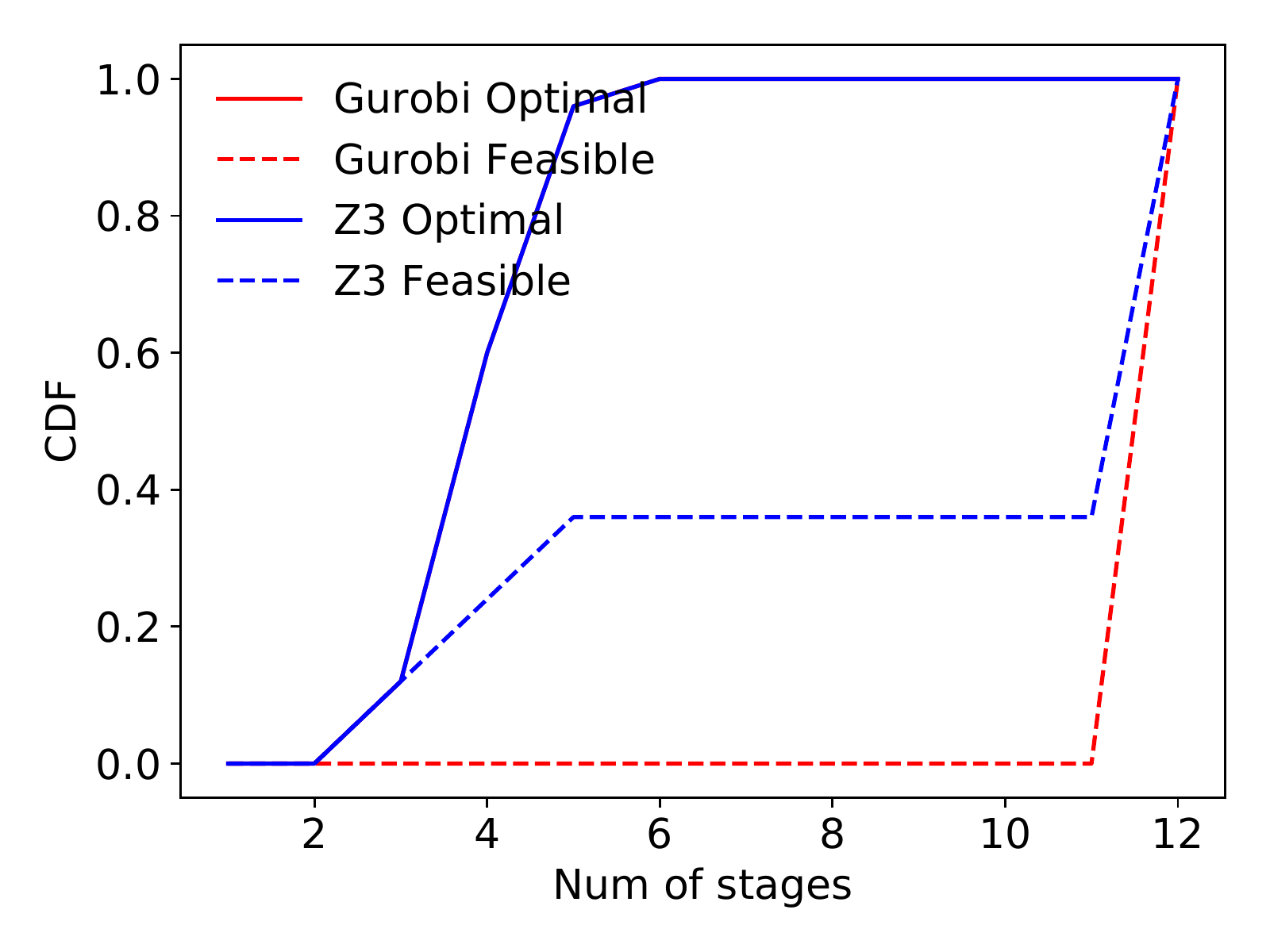}
    \begin{tiny}
    \label{fig:stage_usage}
    \end{tiny}
\end{minipage}
\vspace*{-0.3in}
\caption{Gurobi vs. Z3: Running time, Num of stages.\label{fig:gurobiVsZ3}}
\end{figure*}

\end{document}